\documentclass[twocolumn]{aa} 
\usepackage[]{natbib}
\usepackage{graphicx,color}
\usepackage{txfonts}
\usepackage[colorlinks=true, citecolor=blue]{hyperref}
\usepackage[table]{xcolor}
\usepackage{multirow} 
\usepackage{boldline} 
\usepackage{adjustbox}
\usepackage{float}
\usepackage{booktabs}
\usepackage{caption} 

\newcommand{\orcid}[1]{\href{https://orcid.org/#1}{\includegraphics[width=10pt]{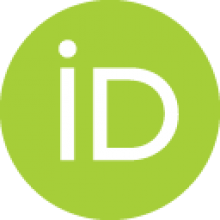}}}

\begin{document} 

\title{Consensus-based Algorithm for Nonparametric Detection of Star Clusters (\textsc{CANDiSC})}

\author{
C.O. Obasi\inst{1}\thanks{Casmir Obasi: casmir.obasi@ucn.cl}
\and J.G. Fern\'andez-Trincado\inst{2,3}\thanks{Jos\'e G. Fern\'andez-Trincado: jose.fernandez@ucn.cl}
\and M. G\'omez\inst{1}
\and D. Minniti\inst{1,4}
\and J. Alonso-Garc\'ia\inst{5,6}
\and B.P. L Ferreira\inst{7}
\and E. R. Garro\inst{8}
\and B. Dias\inst{1}
\and R.K. Saito\inst{9}
\and B. Barbuy\inst{10}
\and M. C. Parisi\inst{11,12}
\and T. Palma\inst{11,13}
\and B.Tang\inst{14}
\and M.Ortigoza-Urdaneta\inst{15}
\and   L. D. Baravalle\inst{12}
\and M.V. Alonso\inst{11,12}
\and F.Mauro\inst{2}
      }

\authorrunning{C.O. Obasi et al.} 
	
\institute{
 Instituto de Astrofísica, Depto. de Fisica y Astronom\'ia, Facultad de Ciencias Exactas, Universidad Andr\'es Bello, Av. Fern\'andez Concha 700, Las Condes, Santiago, Chile
 \and Instituto de Astronom\'ia, Universidad Cat\'olica del Norte, Av. Angamos 0610, Antofagasta, Chile
 \and Universidad Cat\'olica del Norte, Departamento de Ingenier\'ia de Sistemas y Computaci\'on, Av. Angamos 0610, Antofagasta, Chile
\and Vatican Observatory, V00120 Vatican City State, Italy
\and Centro de Astronom\'ia (CITEVA), Universidad de Antofagasta, Av. Angamos 601, Antofagasta, Chile
\and Millennium Institute of Astrophysics (MAS), Nuncio Monse\~nor Sotero Sanz 100, Of. 104, Providencia, Santiago, Chile
\and Departamento de Astronomia, Instituto de Astronomia, Geofísica e Ciências Atmosféricas, Universidade de São Paulo, Rua do Matão 1226, Cidade Universitária, São Paulo 05508-090, Brazil 
\and ESO - European Southern Observatory, Alonso de Cordova 3107, Vitacura, Santiago, Chile
\and Departamento de F\'isica, Universidade Federal de Santa Catarina, Trindade 88040-900, Florian\'opolis, Brazil
\and Universidade de S\~ao Paulo, IAG, Rua do Mat\~ao 1226, Cidade Universit\'aria, S\~ao Paulo 05508-900, Brazil
\and Observatorio Astronómico, Universidad Nacional de Córdoba, Laprida 854, X5000BGR Córdoba, Argentina
\and Instituto de Astronom\'ia Teórica y Experimental (CONICET-UNC), Laprida 854, X5000BGR Córdoba, Argentina
\and Consejo Nacional de Investigaciones Cient\'ificas y T\'ecnicas (CONICET), Godoy Cruz 2290, Ciudad Aut\'onoma de Buenos Aires, Argentina
\and School of Physics and Astronomy, Sun Yat-sen University, Zhuhai 519082, China
\and Departamento de Matem\'atica, Facultad de Ingenier\'ia, Universidad de Atacama, Copiap\'o, Chile
}

	\date{Received ...; Accepted ...}
	\titlerunning{A consensus-based clustering algorithm for VVVX clusters}
	
	
 
  \abstract
   {The VISTA Variables in the Vía Láctea (VVV) and its eXtension (VVVX) are near-infrared surveys mapping the Galactic bulge and adjacent disk. These datasets have enabled the discovery of numerous star clusters obscured by high and spatially variable extinction. However, most previous searches relied on visual inspection of individual tiles, which is inefficient and biased against faint or low-density systems. 
   
   }
   {We aim to develop an automated, homogeneous algorithm for systematic cluster detection across different surveys. Here, we apply our method to VVVX data covering low-latitude regions of the Galactic bulge and disk, affected by extinction and crowding. 
   }
   {We introduce the Consensus-based Algorithm for Nonparametric Detection of Star Clusters (\textsc{CANDiSC}), which integrates kernel density estimation (KDE), Density-Based Spatial Clustering of Applications with Noise (DBSCAN), and nearest-neighbour density estimation (NNDE) within a consensus framework. A stellar overdensity is classified as a candidate if identified by at least two of these methods. We apply \textsc{CANDiSC} to 680 tiles in the VVVX PSF photometric catalogue, covering $\approx$ $1100,\mathrm{deg}^2$. 
}
   {We detect 163 stellar overdensities, of which 118 are known clusters. Cross-matching with recent catalogues yields 5 additional matches, leaving 40 likely new candidates absent from existing compilations. The estimated false-positive rate is below 5\%. 
   }
   {\textsc{CANDiSC} offers a robust and scalable approach for detecting stellar clusters in deep near-infrared surveys, successfully recovering known systems and revealing new candidates in the obscured and crowded regions of the Galactic plane. 
   }

	\maketitle
	
\section{Introduction}
\label{section1}
 Star clusters are essential building blocks of galaxies and serve as key tracers of stellar evolution, Galactic structure, and chemical enrichment histories. In the past two decades, wide-field surveys such as Two Micron All Sky Survey \citep[2MASS;][]{skrutskie2006two}, Wide-field Infrared Survey Explorer \citep[WISE;][]{wright2010wide}, \textit{Gaia} \citep{gaia2016gaia, brown2021gaia}, and the VISTA Variables in the Vía Láctea (VVV) and its extension 
 (VVVX) surveys \citep{minniti2010vista,saito2024vista} have greatly expanded the census of stellar clusters in the Milky Way.

We are now witnessing the arrival of even more powerful wide-field surveys, including Legacy Survey of Space and Time \citep[The LSST collaboration;][]{abell2009lsst}, \textit{Euclid} \citep{blanchard2020euclid}, and the \textit{Nancy Grace Roman Space Telescope} \citep{spergel2015wide}, which promise to deliver unprecedented photometric and astrometric coverage of the Galaxy. These developments have motivated the emergence of automated and semi-automated techniques to identify star clusters in large and complex datasets.

Among the most widely used approaches for star cluster detection are density-based algorithms such as DBSCAN \citep{castro2018new,hunt2023improving}, kinematic methods leveraging \textit{Gaia} proper motions and parallaxes \citep{cantat2018gaia, he2022unveiling}, unsupervised machine learning techniques \citep{castro2020hunting,hao2022newly}, and statistical membership estimation frameworks such as \texttt{UPMASK} and \texttt{pyUPMASK} \citep{krone2014upmask, pera2021pyupmask}. These methods have collectively contributed to the discovery of hundreds of clusters across the Galactic disk, halo, and nearby satellites.

Despite this progress, the inner regions of the Milky Way, particularly the bulge and low-latitude disk, remain relatively unexplored due to severe and spatially variable extinction, as well as high source crowding \citep{minniti2017new,minniti2021survival}. Near-infrared surveys such as the VVVX ( \citealt{saito2024vista}) offer a unique opportunity to probe these obscured regions. However, most cluster searches in the VVV and VVVX data have relied on manual visual inspection of individual tiles \citep[e.g.,][]{bica2018new,minniti2017new,minniti2021discovery,garro2022new,garro2024vvvx}, a process that is time-consuming, subjective, and biased against faint or diffuse clusters.

The primary objective of this work is to develop and apply a homogeneous, fully automated detection algorithm capable of systematically uncovering stellar cluster candidates across the VVVX footprint, with particular sensitivity to those hidden in high-extinction regions. A secondary goal is to minimise the contamination rate in the final candidate list. As argued in \cite{obasi2025multi}, many existing supervised and unsupervised cluster detection pipelines suffer from contamination rates as high as 20--30\%, which can hinder statistical studies of cluster populations.

To this end, we introduce \textsc{CANDiSC}, a consensus-based, unsupervised clustering framework that combines three independent density-based methods to robustly detect stellar overdensities while minimising false positives. In this paper, we present the first application of \textsc{CANDiSC} to the VVVX dataset, demonstrate its ability to recover known clusters, and report the discovery of dozens of new candidate systems. While tailored for the VVV/VVVX survey, \textsc{CANDiSC} is designed to be easily adapted to future large-scale photometric surveys such as LSST, Euclid, and Roman \footnote{\textsc{CANDiSC} is not yet publicly available. A future version incorporating additional multi-survey parameters will be released upon completion of ongoing development.}
.

This paper is structured as follows. Section~\ref{section2} describes the VVVX dataset used in this study. Section~\ref{section3} presents the \textsc{CANDiSC} algorithm and its implementation. Section~\ref{section4} summarises the results of the cluster detection. In Section~\ref{section5}, we discuss the robustness, limitations, and broader implications of our findings. Section~\ref{section6} provides concluding remarks and outlines future directions.

\section{Datasets}
\label{section2}
The VVVX survey \citep{saito2024vista} is the extended phase of the original VISTA Variables in the Vía Láctea (VVV) survey \citep{minniti2010vista}, conducted using the 4.1-m VISTA InfraRed CAMera (VIRCAM) on the VISTA Telescope  \citep{emerson2010visible} at ESO’s Paranal Observatory. The VVVX survey covers approximately 1700 $\mathrm{deg}^2$ in the near-infrared ($J$, $H$, $K_s$) bands, targeting the Galactic bulge and adjacent disk over a longitude range of approximately $-130^\circ < l < +20^\circ$. With its high spatial resolution and 80\% completeness down to $\sim17.5$ mag in the $K_s$ band \citep{saito2024vista}, VVVX is ideally suited for detecting obscured stellar populations in the crowded and highly extincted regions of the inner Galaxy.
\begin{figure*}
    \centering
    \includegraphics[width=1.\linewidth]{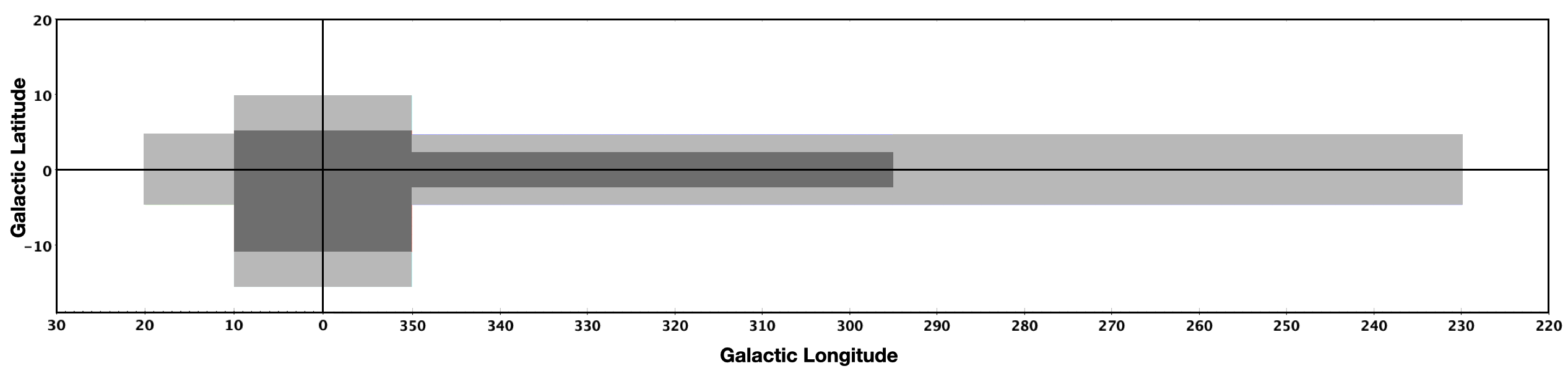}
    \caption{Survey area used in this study is presented. The gray-shaded region indicates the 680 VVVX tiles included in the analysis, while the black-shaded region shows tiles that were excluded.}
    \label{fig:survey_area}
\end{figure*}
The VVVX survey provides deep near-infrared photometry with limiting magnitudes of J $\approx$ 20, H $\approx$ 19, and K$_s$ $\approx$ 17.5, delivered as FITS catalogs containing right ascension (RA), declination (Dec) and magnitudes for point sources. The data used in this study are organized into tiles, each covering 1.646 $\mathrm{deg}^2$ 
and includes 680 VVVX tiles/fields not previously covered in the VVV original footprint. The surveyed region excludes the inner bulge and focuses mainly on the Galactic disk. Figure \ref{fig:survey_area} shows the VVVX footprint used in this work. Tiles that were included in the analysis are shaded in gray, while those that were excluded, particularly near the central bulge, are shaded in black. In this work, we used the point-spread-function (PSF) photometric catalogs produced by Alonso-García et al. (in prep.) using DoPHOT, based on a reprocessing of the VVVX dataset not covered in the VVV original footprint. The catalogs provide high-fidelity photometry for over 700 million sources, with uniform coverage and improved depth compared to previous catalogs based on aperture photometry. 
No photometric cuts or quality filters were applied prior to the application of the \textsc{CANDiSC} algorithm, in order to preserve sensitivity to faint or sparse cluster candidates. The only constraints imposed during detection are a magnitude cut at $K_s < 17.5$ and a color–magnitude filter ($J-K_s$) applied within the algorithm to isolate likely cluster members (see Sect.~\ref{section3}). We caution that some problems may occur in the vicinity of very bright saturated objects.

\section{Methodology}
\label{section3}
\subsection{Overview}

In this section, we present the Consensus-based Algorithm for Nonparametric Detection of Star Clusters (\textsc{CANDiSC}), developed to identify stellar overdensities in the crowded and highly extincted Galactic bulge and disk using near-infrared data from the VVVX survey \citep{saito2024vista}. \textsc{CANDiSC} combines three unsupervised density estimation techniques commonly applied in the literature for overdensity detection: Gaussian Kernel Density Estimation (KDE) \citep{Parzen1962}, Density-Based Spatial Clustering of Applications with Noise (DBSCAN) \citep{ester1996density}, and Nearest-Neighbor Density Estimation (NNDE) \citep{loftsgaarden1965nonparametric}. To enhance robustness against noise and variable cluster morphologies, \textsc{CANDiSC} employs a consensus voting scheme to select candidate cluster members. This ensemble approach is particularly well suited for detecting faint or irregular clusters in crowded fields, such as the inner bulge and disk. By combining the outputs of multiple unsupervised methods, the algorithm increases robustness against false positives and enhances sensitivity to low-density overdensities that may be missed by individual classifiers. As demonstrated in this work, the code successfully identifies candidate clusters with a few members and recovers systems with non-symmetric or sparse morphologies.

\subsection{Data Preprocessing}\label{data_processing}
To reduce foreground contamination and enhance contrast in the highly reddened Galactic bulge and disk, we apply a color–magnitude filter:
\begin{equation}(J - K_s) \in [0.4, 1.4], \quad J < 18.5, \quad K_s < 17.5.,\end{equation}
designed to retain distant main-sequence and giant stars while minimizing the contribution from nearby dwarfs and background galaxies. These cuts are optimized for the depth and extinction characteristics of the VVVX survey in the inner bulge area (A$_{K_s}$ $\approx$ 0.24–2.5 mag).
\begin{figure}
    \centering
    \includegraphics[width=1.\linewidth]{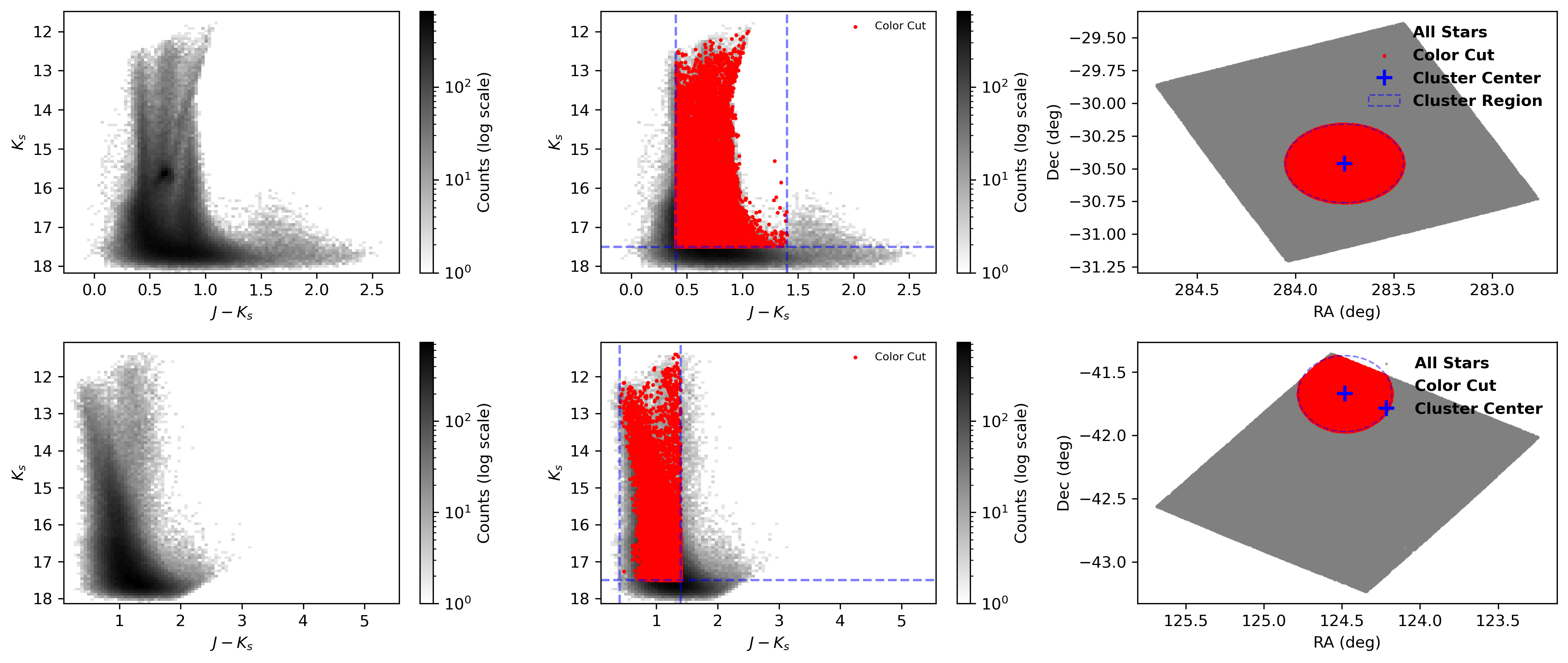}
    \caption{Color-magnitude and spatial distributions of stars in tiles \texttt{b0411} (M54) upper panel and \texttt{e0621} (Pismis 2) bottom panel. Left panels: 2D histograms showing all sources in the $(J - K_s)$ vs. $K_s$ diagram. Middle panels: Sources passing the color-magnitude selection ($0.4 < J - K_s < 1.4$, $K_s < 17.5$) are overlaid in red, with the selection boundaries marked by dashed blue lines. Right panels: Spatial distributions in RA vs. Dec, where all stars are shown in gray, and color-selected sources in red. Blue crosses mark the cluster centers, and dashed blue circles indicate a $0.3^\circ$ radius.}
    \label{fig:color_cut}
\end{figure}
Figure~\ref{fig:color_cut} presents color-magnitude and spatial diagnostic plots for two known clusters in our sample: the globular cluster M54 (tile \texttt{b0411}, centered at RA = 283.75$^\circ$, Dec = -30.46$^\circ$), and the open cluster Pismis 2 (tile \texttt{e0621}, centered at RA = 124.48$^\circ$, Dec = -41.67$^\circ$). Each row in the figure corresponds to one cluster: M54 in the top panels, and Pismis 2 in the bottom panels. The left panel of each row  
displays $(J - K_s)$  vs $K_s$ color magnitude diagram of all the stars in the tile/field where the cluster center is located. The log-normalized color scale indicates stellar density, revealing a broad distribution peaking around $J - K_s \sim 0.5$-1.5 and $K_s \sim 14$-17, consistent with a mix of cluster members and field stars. The middle panels highlight stars selected using the color-magnitude cuts described above, overlaid in red on the same CMDs. The blue dashed lines indicate the filter boundaries. This selection effectively isolates the cluster's main sequence and red giant branch, although some field contamination remains. The right panels show the spatial distribution of stars in equatorial coordinates. All stars are shown in gray, while color-cut selected stars are plotted in red. A blue cross marks the cluster center, and a dashed blue circle denotes a 0.3-degree radius region used to visualize the cluster extent.

For computational efficiency, we implement a conditional downsampling scheme for large catalogues: we retain 10\% of sources for datasets (each photometric tile processed) exceeding 3 million entries, 30\% for those above 1.7 million, 60\% for $ >$1.4 million, and retain all sources for catalogues smaller than 1.2 million. Random selection ensures spatial uniformity while significantly reducing memory usage and maintaining statistical completeness.

\subsection{Spatial Filtering and Subtiling}

To focus on high-density regions, we apply spatial binning using a 2D histogram of RA and Dec with 120 bins per degree, resulting in bin sizes of approximately 0.00833 degrees (30 arcsec) in both RA and Dec, scaled to the data's extent.  Only bins with source counts above the 80th percentile of the non-empty bin distribution are retained. To prevent overfitting in crowded fields, bins with more than 10,000 sources are further downsampled to a maximum of 10,000 sources by random selection. The filtered data are subdivided into a $4 \times 4$  grid of spatial subtiles to account for localized extinction and density gradients. Subtiles with fewer than 5 sources are excluded to ensure robust clustering.

\subsection{Density Estimation Techniques}
\textsc{CANDiSC} employs three unsupervised methods to identify overdense regions within each subtile, using sigma-clipped statistics (3$\sigma$ clipping) to define robust local density thresholds. In the following subsection, we describe each method in detail and summarise how \textsc{CANDiSC} combines their outputs in a consensus approach to flag a target as a cluster candidate.

\subsubsection{Kernel Density Estimation (KDE)}
KDE is a widely used non-parametric method for estimating the continuous spatial density of point sources, particularly effective in crowded stellar fields \citep[][and references therein]{vio1994probability,ferdosi2011comparison,seleznev2016open,nambiar2019star}. It works by placing a Gaussian kernel on each star's position and summing the contributions to produce a density estimate at each star's location. In our implementation, KDE uses a Gaussian kernel with a bandwidth of $h$, optimized for the typical angular sizes of VVVX clusters (\(\sim0.05-0.2^\circ\)). After testing several values, we found \(h = 0.1^\circ\) to be most effective. The local density at a star's position \(\mathbf{x}\) is estimated as:
\begin{equation}
\rho(\mathbf{x}) = \frac{1}{n h^2} \sum_{i=1}^{n} \exp \left( -\frac{||\mathbf{x} - \mathbf{x}_i||^2}{2h^2} \right),
\end{equation}
where \(n\) is the number of stars and \(\mathbf{x}_i\) are their positions. The choice of \(h\) is critical: smaller values retain compact structures but amplify noise, while larger values suppress noise but may oversmooth true overdensities \citep[see e.g.,][]{ferdosi2011comparison}. To identify candidate clusters, we select stars with KDE-estimated densities \(\rho\) exceeding a threshold of \(\bar{\rho} + \sigma\), where \(\bar{\rho}\) and \(\sigma\) are the \(\sigma\)-clipped mean and standard deviation (using a 1\(\sigma\) clip) of the density field, reducing the impact of outliers and background inhomogeneities. Stars with \(\rho > \bar{\rho} + 5\sigma\) are flagged as overdense, effectively isolating smooth, centrally concentrated stellar systems.

\subsubsection{DBSCAN}
DBSCAN is a widely used unsupervised clustering algorithm that identifies dense groupings of points without requiring assumptions about cluster number or shape \citep[see e.g.,][]{castro2020hunting,he2022unveiling,he2022blind,prisinzano2022low,strantzalis2024robust}. Clusters are defined as regions where a minimum number of stars (min\_samples) lie within a specified neighborhood radius ($\varepsilon$). Points satisfying this condition are labeled as core points, while nearby points are border points. Isolated points that do not meet either criterion are classified as noise.
In our implementation, we adopt $\varepsilon = 0.1^\circ$ and min\_samples = 5, values optimized for typical VVVX cluster sizes. For each group identified by DBSCAN, we compute its effective density as:
\begin{equation}
\rho = \frac{N}{\pi \varepsilon^2},
\end{equation}
where $N$ is the number of member stars.
We define overdense clusters as those with $\rho > \bar{\rho} + 5\sigma$, where $\bar{\rho}$ and $\sigma$ are the mean and standard deviation of cluster densities after 3$\sigma$ clipping, ensuring robustness against outliers. DBSCAN’s ability to recover arbitrarily shaped structures makes it well suited for the crowded and highly variable stellar environments of the inner Galactic bulge. Moreover, it provides a complementary detection strategy to KDE, enhancing redundancy and robustness in our overall pipeline, with both methods using 3$\sigma$ clipping for consistency in outlier removal (though KDE applies a less stringent initial threshold of $\bar{\rho} + \sigma$ for candidate selection).

\subsubsection{Nearest-Neighbor Density Estimation (NNDE)}

The third clustering algorithm employed in this study is NNDE \citep{loftsgaarden1965nonparametric}, a non-parametric method that estimates local stellar density based on the distance to the $k$-th nearest neighbor. The density around each star is given by:
\begin{equation}
  \rho = \frac{k}{\pi d_k^2},
 \end{equation}
where $d_k$ is the distance to the $k$-th nearest neighbor. 
In our implementation, we adopt \(k = 5\), based on sensitivity tests using values of \(k = 3, 5, 7,\) and \(10\), which showed that \(k = 5\) offers a good balance between sensitivity to local overdensities and robustness against noise in the VVVX dataset. This choice is also supported by statistical studies recommending small \(k\) values for local density estimation \citep[e.g.;][]{kung2012optimal}.
We identify overdense sources as those with $\rho > \bar{\rho} + 5\sigma$, where $\bar{\rho}$ and $\sigma$ are calculated after applying 3$\sigma$ clipping to the full density distribution. NNDE is highly sensitive to local variations in source density and naturally adapts to changes in crowding, making it effective in both the dense stellar fields of the Galactic bulge and more diffuse regions \citep{casertano1985core}. NNDE serves as the final redundancy to KDE and DBSCAN, enhancing the algorithm’s ability to detect low-contrast and irregularly shaped stellar overdensities that may be missed by the other two methods. 
\begin{figure}
    \centering
    \includegraphics[width=0.5\linewidth]{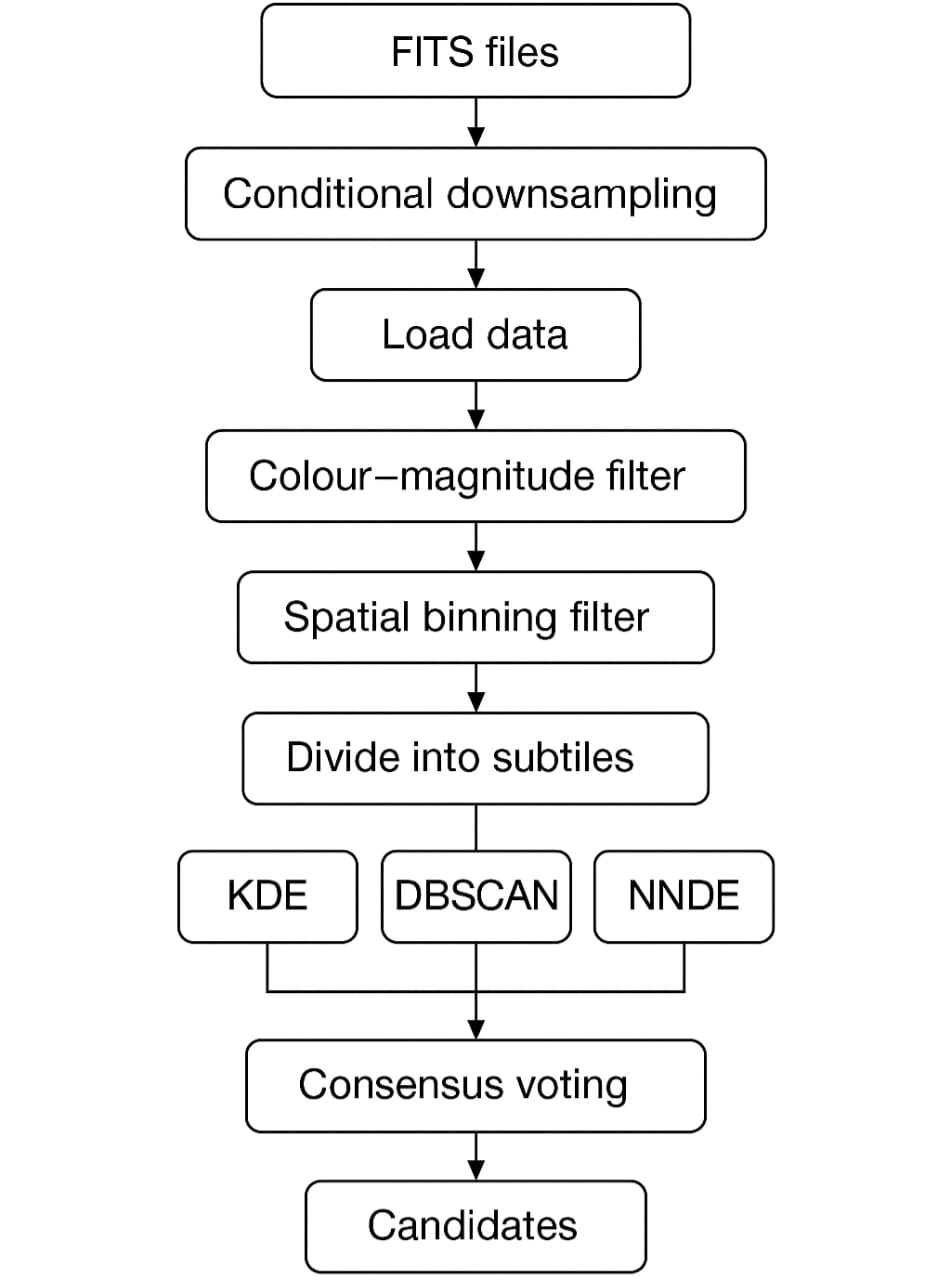}
    \caption{Flowchart of \textsc{CANDiSC}, the Consensus-based Algorithm for Nonparametric Detection of Star Clusters.}
    \label{fig:flowchart}
\end{figure}

\subsection{Consensus Detection Strategy}

Each of the three detection techniques: KDE, DBSCAN, and NNDE produces a binary mask indicating whether a star lies in a locally overdense region within a given subtile. To reduce method-specific biases and improve detection reliability, we implement a consensus voting scheme. A star is flagged as a candidate cluster member if it is identified as overdense by at least two of the three methods:
 \begin{equation}
 \text{Candidate}_i = \left( \sum_{j=1}^{3} Dec_{ij} \right) \geq 2,
\end{equation}
 where $Dec_{ij} = 1$ if method $j$ identifies star $i$ as overdense, and 0 otherwise.
 This majority-voting approach minimizes spurious detections caused by noise or local field fluctuations in any single method while maintaining sensitivity to genuine stellar overdensities. By combining independent density estimators, the final list of candidate members is statistically robust and physically meaningful, especially important in the crowded and differentially extinct fields. A schematic overview of the detection pipeline is shown in Figure~\ref{fig:flowchart}.

\subsection{Parallel Execution and Output}
Given the high computational demand, especially for large tiles, the pipeline is parallelized using \texttt{joblib.Parallel} with memory-aware core allocation. 

 Candidate stars (RA, Dec) are saved as FITS tables for further analysis. The pipeline is implemented in Python, using scikit-learn \citep{pedregosa2011scikit} for DBSCAN and NNDE, scipy \citep{virtanen2020scipy} for KDE, and astropy \citep{robitaille2013astropy} for FITS handling and sigma-clipped statistics.
 
\subsection{Validation of the Algorithm with VVVX data}
\label{sec:validation}
To validate the performance and reliability of {\textsc{CANDiSC}, we applied the full detection pipeline to a test sample of well-characterised globular and open clusters within the footprint of the VVVX survey. These clusters are distributed across varying degrees of crowding, extinction, and structural concentration, providing a representative benchmark for inner Galaxy cluster detection. The sample, summarised in Table~\ref{tableA}, includes classical bulge globular clusters such as M54, NGC 6652, NGC 6293, and NGC 6325, as well as looser open clusters like ESO 425-3. The columns show the VVVX tile name, literature cluster name, object type, central coordinates (J2000), and the number of member stars detected by \textsc{CANDiSC}.

\texttt{\textsc{CANDiSC}} was executed on ten VVVX tiles containing these clusters, each processed through the full pipeline described in Section~\ref{section3}. For each tile, the algorithm outputs a FITS file named \texttt{membership\_candidates.fits}, which stores the RA and Dec of all stars identified as candidate cluster members.

The detection parameters for each of the underlying methods, KDE bandwidth $h$, DBSCAN neighborhood radius $\varepsilon$ , and \texttt{min\_samples}, NNDE neighbor count $k$, and the overdensity detection threshold $\sigma$, were optimized through an extensive grid search combined with objective performance evaluation test. For KDE, $h$ was varied between 0.05$^\circ$ and 0.2$^\circ$ in steps of 0.025$^\circ$, which reflects the expected angular sizes of VVVX clusters. For DBSCAN,  $\varepsilon$  ranged from 0.05$^\circ$ to 0.15$^\circ$ in steps of 0.025$^\circ$, and \texttt{min\_samples} from 3 to 10. For NNDE, $k$ values of 3, 5, and 7 were tested. The detection threshold $\sigma$ was varied between 3.0 and 6.0. We selected optimal parameter combinations by maximizing the recovery rate of known cluster members and minimizing false positives, as quantified through cross-matching with the literature catalog of \cite{kharchenko2013global}. Finally, visual inspection of the output density maps was conducted, and it serves as an additional qualitative check to confirm spatial clustering.

Each detection was then manually inspected to verify the presence of spatial clustering and its consistency with known cluster positions. Remarkably, even in cases of highly dispersed open clusters such as ESO~425$-$3 (detected in tile~\texttt{e0609}), a single overdensity was sufficient to justify re-running the code with an adjusted set of parameters particularly by widening/contracting the range of the colour cut, as described below. When the object is statistically significant, modifying certain parameters can enhance the number of detected members, thereby improving the overall recovery of the cluster members.


Figures~\ref{fig:valid1}-\ref{fig:valid2} illustrate the validation results. The left panels show the stellar density distribution in fields containing M54 (b0411), NGC 6652 (b0436), NGC 6293 (b0490), NGC 6325 (b0492), CWNU 4193 (e0618), CL Pismis 2 (e0621), CL Haffner 15 (e0613), and M19 (b0503). The right panels show the same fields with overplotted candidate members identified by \textsc{CANDiSC}.

Positional accuracy of the detected candidates exceeds 99\%, as confirmed by cross-matching with cluster centers from the literature. The algorithm is highly selective, identifying only localized overdensities consistent with known clusters, and avoids false positives in uniform fields. In Section \ref{false_postive}, we report that the overall false positive rate is below 5\%, with the majority of spurious detections attributable to dark nebula, variable star and eclipsing binaries in crowded fields, likely flagged due to local density fluctuations or photometric outliers. Notably, each of these spurious detections contains fewer than five member stars. 
We note that for the sparse/open clusters, expanding the color cut to 0.3$\leq$J-K${_s}$ $\leq$ 1.4 increases the number of detected stars significantly (e.g ESO 425-3, which was improved from 1 to 15 members.) This suggests the need for a flexible color cut in low-density environments.

These results affirm that \textsc{CANDiSC} reliably recovers genuine stellar overdensities in both compact and diffuse clusters, and can operate effectively in the challenging inner Galaxy environment targeted by infrared surveys like VVVX.
\begin{table*}[ht!]
\caption{List of clusters used for the validation of the \textsc{CANDiSC} code.}
\label{tableA}
\centering
\scriptsize
\setlength{\tabcolsep}{3pt} 
\begin{tabular}{@{}c l l c c cc r@{}}
\hline
Tile Name & Literature Name & Obj$_{\text{type}}$ & RA & Dec & No. of stars recovered (0.4$\leq$ J-K$_s$ $\leq$ 1.4) & No. of stars recovered (0.3$\leq$ J-K$_s$ $\leq$ 1.4)\\

\hline
e0609  & ESO 425-3     & OpC & 113.87 & -27.70 & 1 & 15    \\
b0411  & M 54          & GC  & 283.75 & -30.46 & 1,084&1019 \\
b0436  & NGC 6652      & GC  & 278.95 & -32.98 & 310  &348 \\
b0461  & NGC 6316      & GC  & 259.15 & -28.14 & 12 & 37   \\
b0490  & NGC 6293      & GC  & 257.54 & -26.57 & 86 &137   \\
b0492  & NGC 6325      & GC  & 259.49 & -23.76 & 152&108   \\
b0503  & M 19          & GC  & 255.66 & -26.28 & 71&-    \\
e0613  & Cl Haffner 15 & OpC & 116.40 & -32.83 & 47&55    \\
e0618  & CWNU4193      & GC  & 121.17 & -38.57 & 19&15    \\
e0621  & Cl Pismis 2   & OpC & 124.48 & -41.67 & 32&328    \\
\hline
\end{tabular}
\end{table*}

\begin{figure*}
    \centering
    \includegraphics[width=0.45\linewidth]{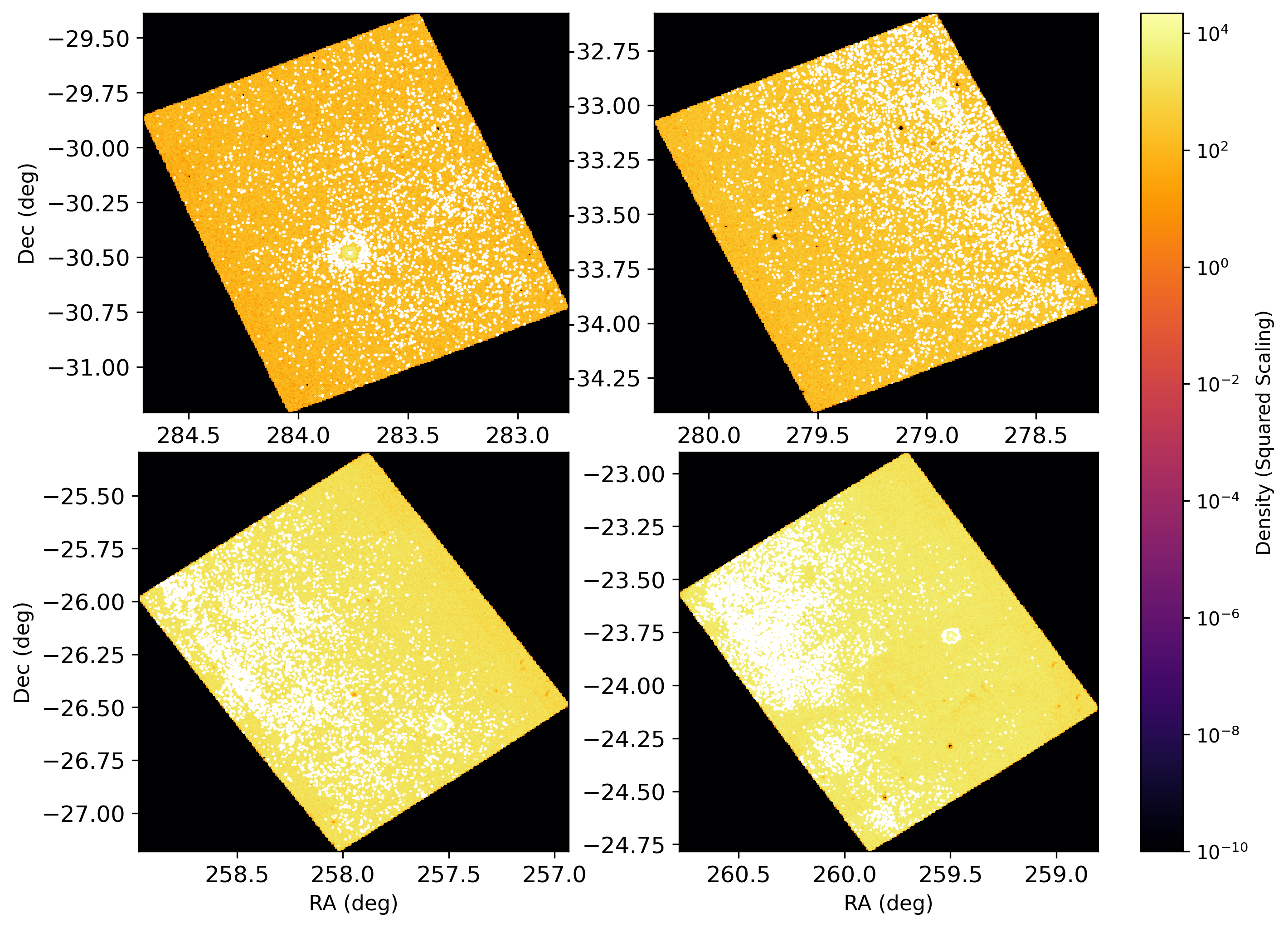}
   \includegraphics[width=0.45\linewidth]{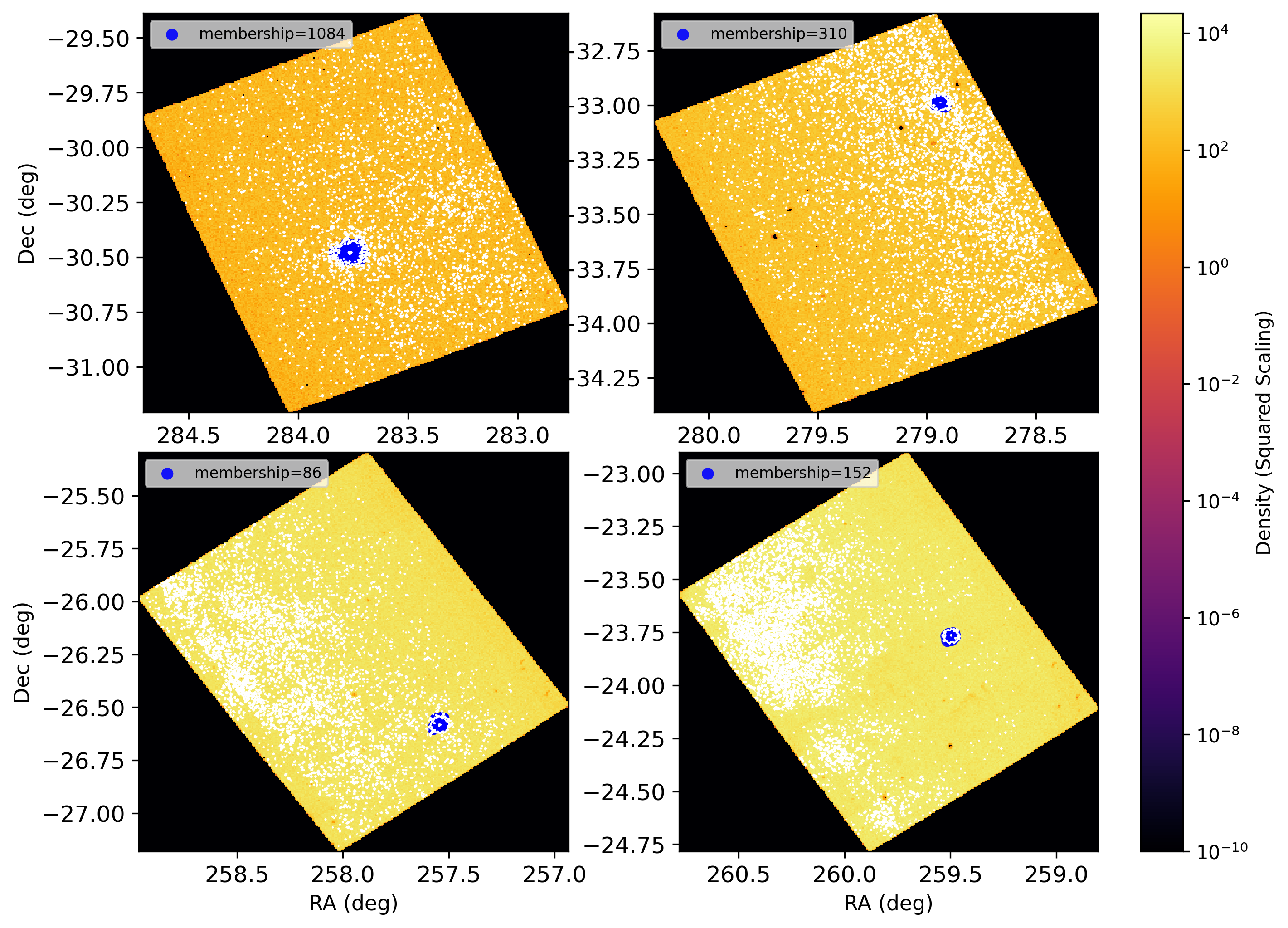}
    
    \caption{Density distribution maps for the VVVX tiles containing clusters used to validate the \textsc{CANDiSC} code. \textit{Left panel:} Stellar density maps for the fields of M54 (b0411), NGC 6652 (b0436), NGC 6293 (b0490), and NGC6325 (b0492). The upper and lower subpanels correspond to different tiles. \textit{Right panel:} Same maps as in the left panel, now overplotted with the candidate cluster members identified by \textsc{CANDiSC}. The legend indicates the number of recovered members for each cluster.
}
    \label{fig:valid1}
\end{figure*}
\begin{figure*}
    \centering
    \includegraphics[width=0.45\linewidth]{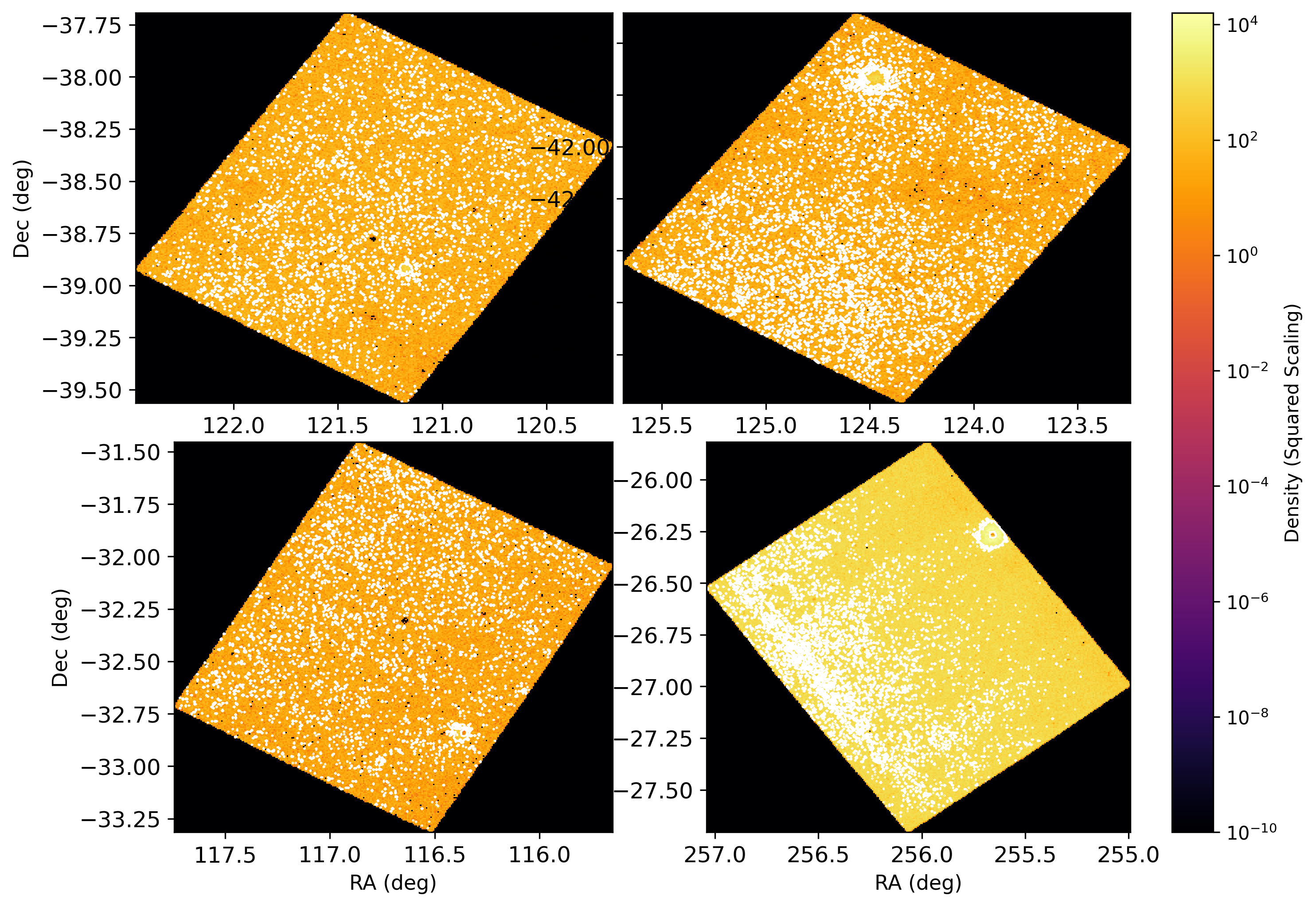}
   \includegraphics[width=0.45\linewidth]{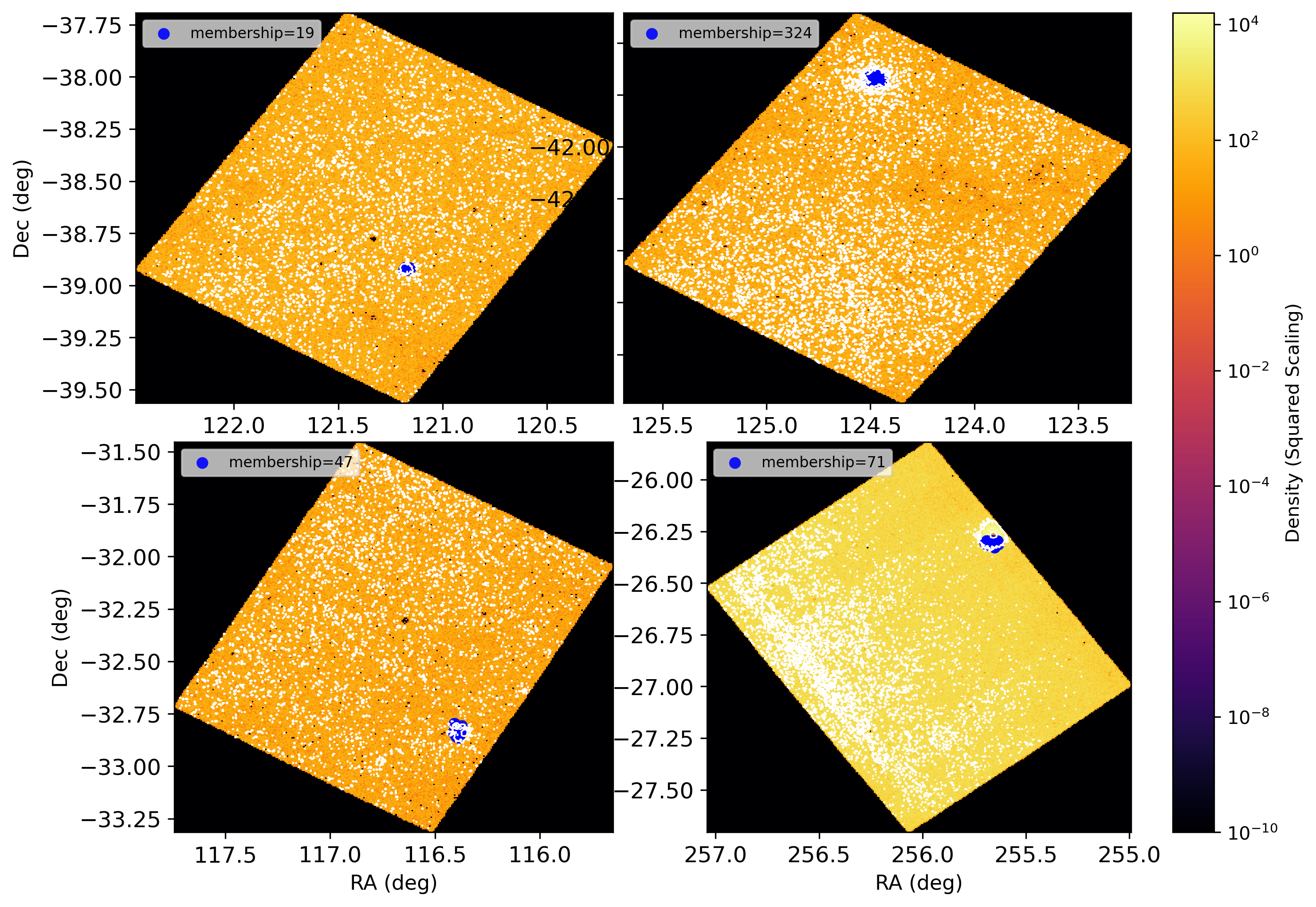}
    
    \caption{Same as in Fig.~4, but for CWNU 4193 (e0618), Pismis 2 (e0621), Haffner 15 (e0613), and M 19 (e0503). The maps show the stellar density distributions in each VVVX tile, with the identified cluster members overplotted. The legend reports the number of recovered members for each cluster.
}
    \label{fig:valid2}
\end{figure*}

\subsection{Validation of the Algorithm with Synthetic Data}
\begin{figure}
    \centering
    \includegraphics[width=0.7\linewidth]{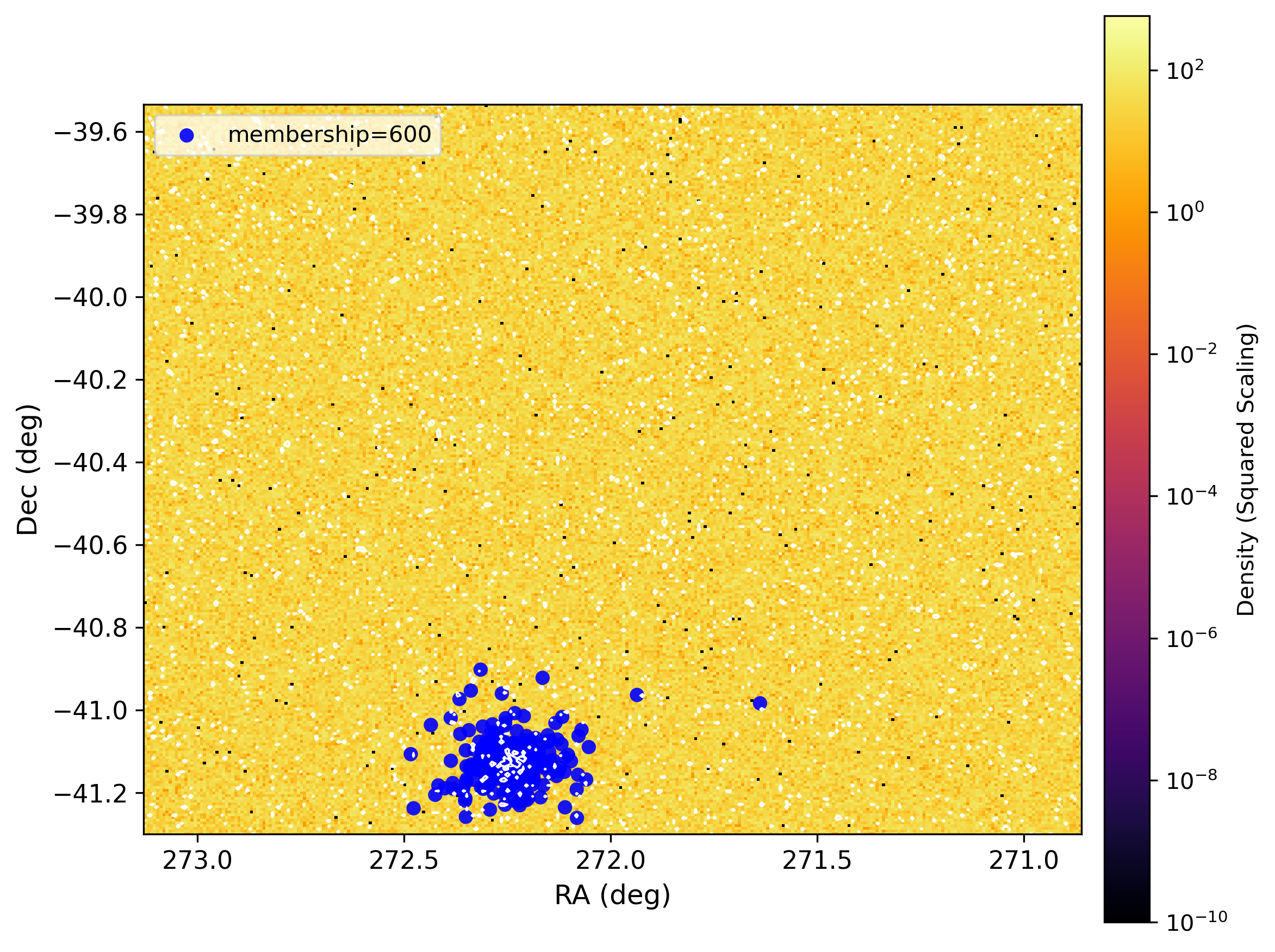}
    \caption{Density distribution map of the synthetic injection field in RA, DEC. The background color scale represents the squared and logarithmically normalized spatial density of all injected stars. 
    Blue points denote the positions of the recovered members of the injected cluster containing 600 stars.}
    \label{fig:validationdistr}
\end{figure}
\begin{figure*}
    \centering
    \includegraphics[width=0.5\linewidth]{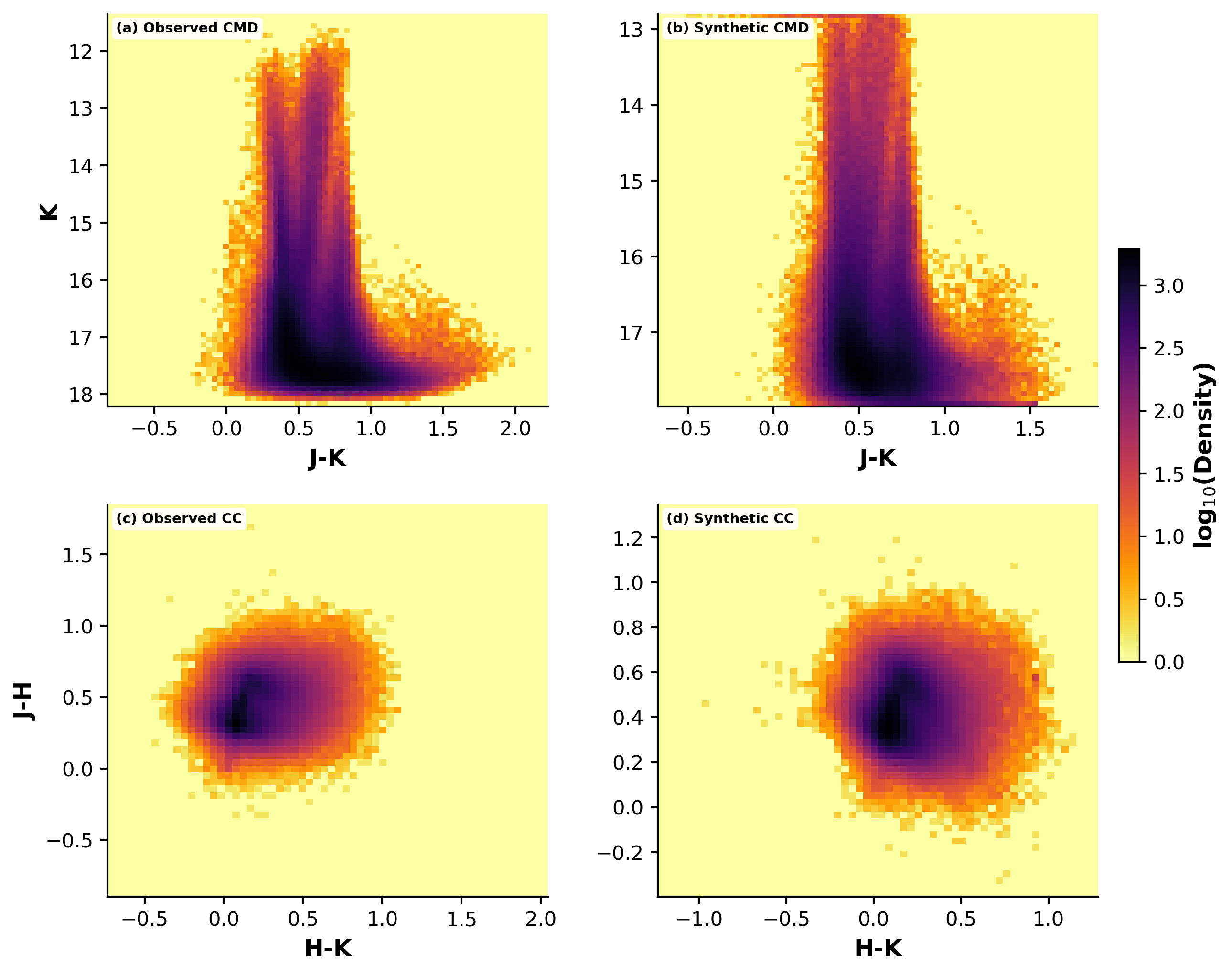}
    \caption{Observed (left) and synthetic (right) of the VVVX-like color-magnitude (top panels) and color-color diagram (bottom panels). The synthetic catalog reproduces the observed near-infrared stellar locus and photometric distributions using KDE resampling and a locus prior.}
    \label{fig:syntheticCmD}
\end{figure*}
\begin{figure*}
    \centering
    \includegraphics[width=0.4\linewidth]{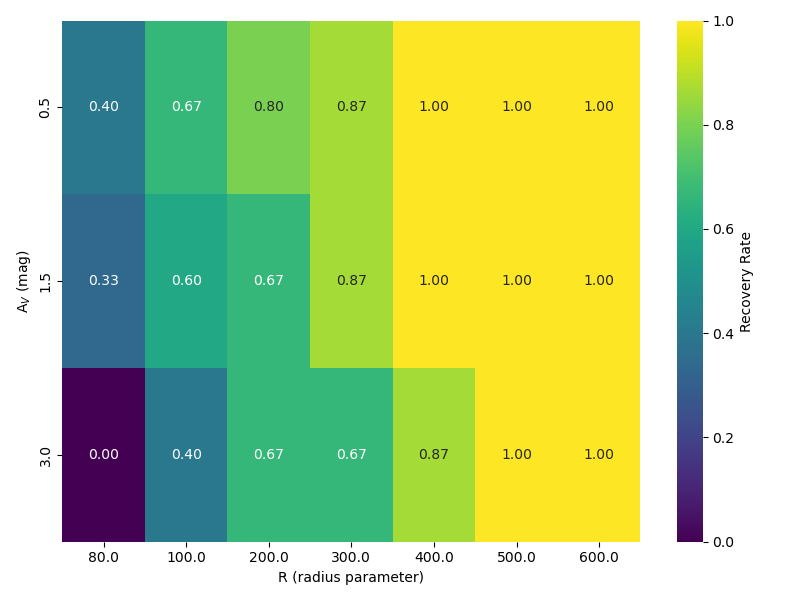}
    \includegraphics[width=0.4\linewidth]{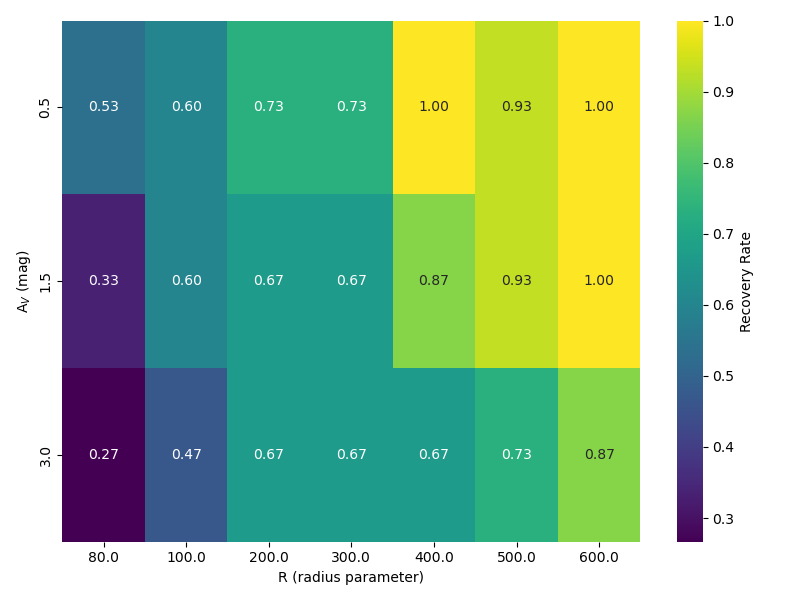}
    \caption{Two-dimensional completeness surface (heatmap) showing the mean recovery rate as a function of cluster radius ($R$) and extinction ($A_V$) (left-right panel default and tuned configuration). For the default configuration, completeness decreases toward smaller radii and higher extinction, and becomes extreme at R=80, where it drops to 0. In contrast, for tuned settings, completeness gradually increases toward larger radii and lower extinction. The plot also shows the parameter space where the pipeline remains reliable.}
    \label{fig:completeness_surface}
\end{figure*}
We constructed a synthetic dataset designed to reproduce the characteristics of the VVVX observations, allowing us to quantify in a controlled setting the detection efficiency and photometric recovery performance of our pipeline, as well as to expose potential limitations that cannot be directly probed with real data. A suite of artificial stellar clusters was injected into realistic VVVX fields as shown in figure~ \ref{fig:validationdistr}, where we show one example of our generated synthetic data overplotted with the injected cluster members. These simulations were generated using a custom \texttt{Python} framework that models both field and cluster populations via kernel density estimation (KDE) of the observed photometric distributions.

Figure~\ref{fig:syntheticCmD} illustrates an example of the observed and synthetic datasets (left–right), with the upper panels showing their respective CMDs and lower panels their colour–colour diagram distributions. Field stars were resampled from the empirical J-K$_s$ vs K$_s$ and J-H vs H-$K_s$ density distributions of the input catalog. This ensures that the synthetic background accurately reproduces the color–magnitude structure and completeness characteristics of the data.

Cluster members were spatially distributed according to a Plummer-like profile \citep{plummer1911problem} and assigned photometry drawn from the same KDE models, with an optional locus prior to preserve the median (J-$K_s$ vs $K_s$) relation of the field population. Differential reddening was simulated by applying a Gaussian extinction patch centered on each cluster, adopting extinction ratios $A_J/A_V = 0.282$, $A_H/A_V = 0.175$, and $A_{K_s}/A_V = 0.114$ \citep{cardelli1989relationship}.

The simulations span a three-dimensional grid of cluster parameters: richness ($N_\star = 80$–$600$), angular half-light radius ($r_h = 0.005$–$0.03^\circ$), and line-of-sight extinction ($A_V = 0.5$–$3.0$~mag). Each configuration was realized five times to account for stochastic variance, resulting in a total of 315 synthetic fields. Every realization is stored as a combined \texttt{FITS} catalog containing both field and cluster stars, together with a separate \texttt{CSV} catalog listing only cluster members. This dataset provides a reproducible and statistically controlled framework for evaluating the completeness and parameter-recovery accuracy of our cluster detection algorithm.

After creating our synthetic dataset, we processed it through the \textsc{CANDiSC} pipeline using the default settings described in Section~\ref{sec:validation}. We initially found that the pipeline has difficulty recovering clusters with fewer than 60 members, as these tend to blend with the field stars. Based on this, we set a minimum cluster size of 80 members and then explored the effect of varying key parameters to enhance detection sensitivity. Specifically, we adjusted the spatial binning threshold to the 90th percentile of non-zero histogram bins (from the default 80th percentile) to focus on denser regions. For overdensity detection, we set the KDE bandwidth to 0.09 (from 0.1), the DBSCAN $\epsilon$ to 0.08 (from 0.1) with a minimum of 15 samples (from 5), and the NNDE $k$ to 10 (from 5), while increasing the sigma threshold to 4.0 (from 3.0) for all methods to enforce stricter overdensity criteria. The results and detailed diagnostics of these analyses are presented in Appendix~\ref{AppendixB}. Here, we summarise only the global completeness behavior relevant to the main catalog. A full evaluation of the recovery distributions, the dependence on extinction and cluster size, the astrometric offsets, and the purity–completeness relation is provided in Appendix \ref{AppendixB}.
To summarise the pipeline’s performance across the full parameter space of the synthetic injections, we present in Figure~\ref{fig:completeness_surface} the completeness surface maps for the default and tuned configurations. These maps highlight the global dependence of detection efficiency on cluster richness and extinction, and they provide the overview that complements the detailed diagnostics discussed in Appendix~\ref{AppendixB}. The default configuration (left panel) shows a clear “sweet spot’’ at $R \sim 300$ and $A_V \sim 1$~mag, where completeness exceeds 80\%, with detection limits primarily driven by extinction and cluster size. The tuned configuration (right panel) shows notable improvements at the low-richness and high-extinction edges of the parameter space. For example, the recovery rate increases to 0.53 at $A_V = 0.5$, $R = 80$, and to 0.27 at $A_V = 3.0$, $R = 80$, compared to 0.40 and 0.00 in the default case. While these gains indicate enhanced sensitivity to faint or reddened clusters, the improvement across intermediate richness values is less uniform, with the most significant enhancement occurring near the lower $R$ boundary.

\section{Results}
\label{section4}
We applied the \textsc{CANDiSC} algorithm to the full set of VVVX tiles/fields not previously covered in the VVV original footprint. Each tile corresponds to a specific region in the VVVX survey, and in total, we analyzed 680 tiles. After processing, \textsc{CANDiSC} identified 163 candidate stellar overdensities.
Among these, 118 objects correspond to known entries in the SIMBAD database\footnote{\url{https://simbad.cds.unistra.fr/simbad/sim-fcoo}}, including several clusters previously discovered in the VVVX footprint \citep[see][]{garro2022unveiling, garro2024over}. Each of these candidate stellar overdensities was visually inspected using composite images available in SIMBAD, including Pan-STARRS DR1 color images (constructed from $g$ and $z$ bands) and DECam Plane Survey DR1 images (in $g$, $r$, and $z$ bands). These visual inspections confirmed that the spatial overdensities are consistent with stellar clusters. Representative examples of the newly detected candidates are shown in Figure~\ref{fig:composite}. In the upper panels, we show cluster candidates detected in tiles \texttt{e0602} (left) and \texttt{e1022} (right), located at $RA,Dec = 108.095^\circ,\ -18.160^\circ$ and $127.91^\circ,\ -41.78^\circ$, respectively. The lower panels display candidates from tiles \texttt{e0965} (left) and \texttt{e1047} (right), centered at $RA,Dec = 275.12^\circ,\ -14.23^\circ$ and $110.75 ^\circ,\ -16.67^\circ$, respectively.

\begin{figure}
    \centering
    \includegraphics[width=0.35\linewidth]{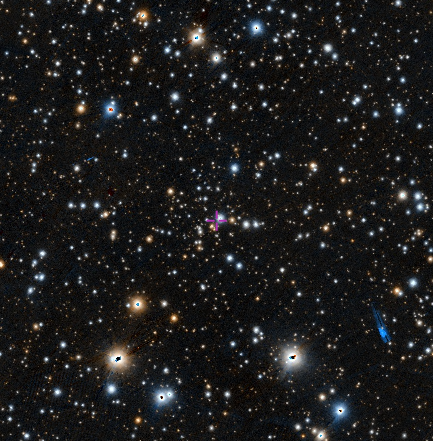}
    \includegraphics[width=0.35\linewidth]{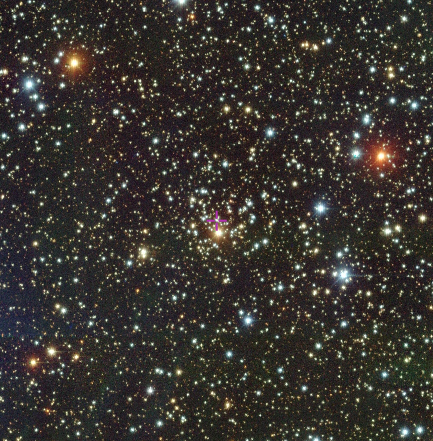}\\
    
    \includegraphics[width=0.35\linewidth]{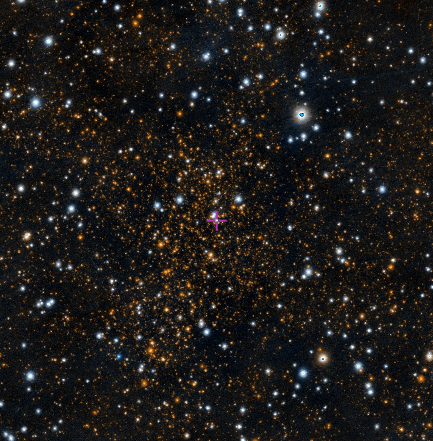}
    \includegraphics[width=0.35\linewidth]{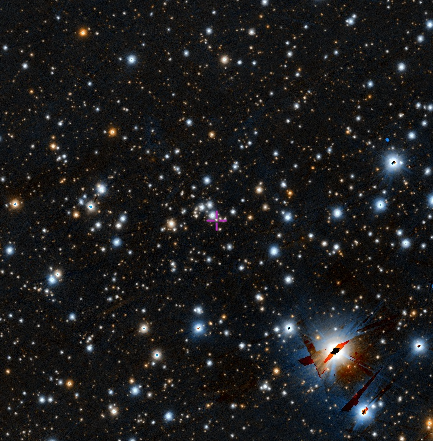}
    
    \caption{Composite images of a subset of newly detected cluster candidates are shown. The upper row displays clusters from tiles e0602 and e1022, while the lower row shows clusters from tiles e0965 and e1047. Panels (a) and (c) are based on Pan-STARRS images, and panels (b) and (d) use DECam Plane Survey (DECaPS) DR1 color composites constructed from the g, r, and z bands.
}
    \label{fig:composite}
\end{figure}
\begin{figure}
    \centering
    \includegraphics[width=1\linewidth]{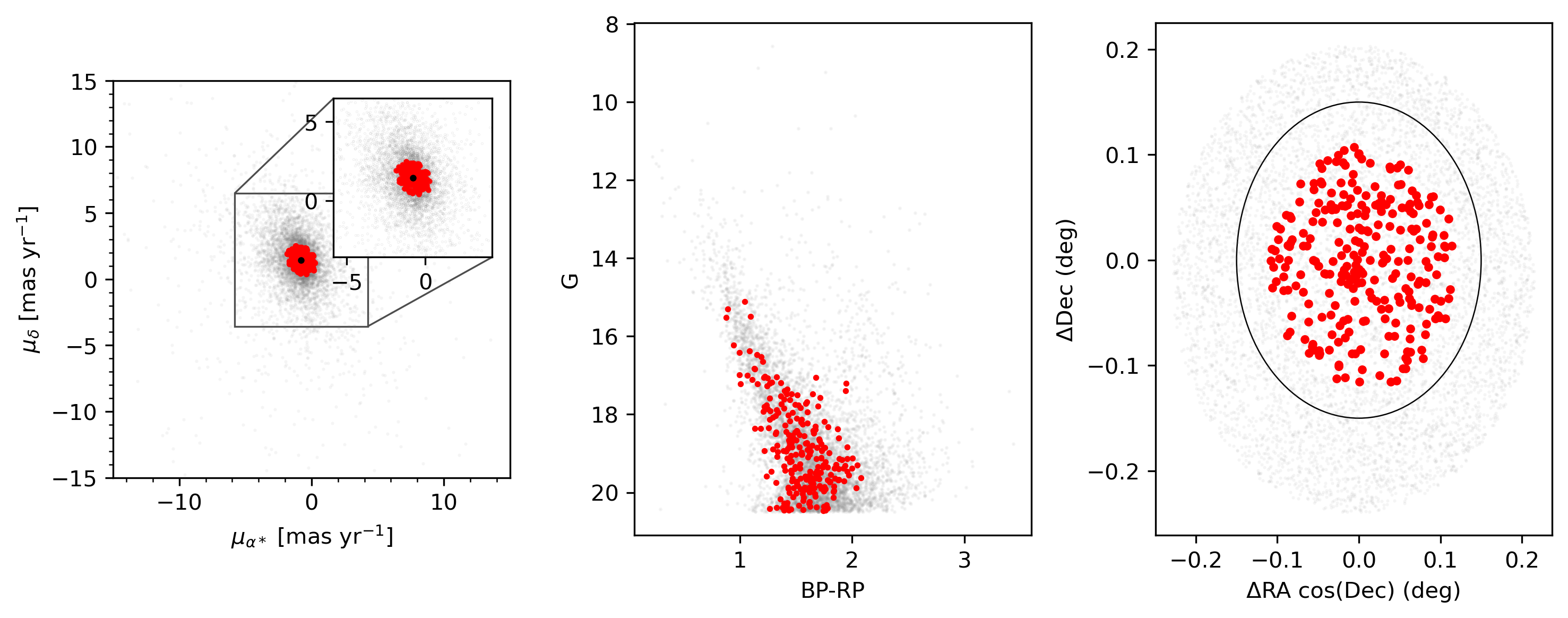}
    \includegraphics[width=1\linewidth]{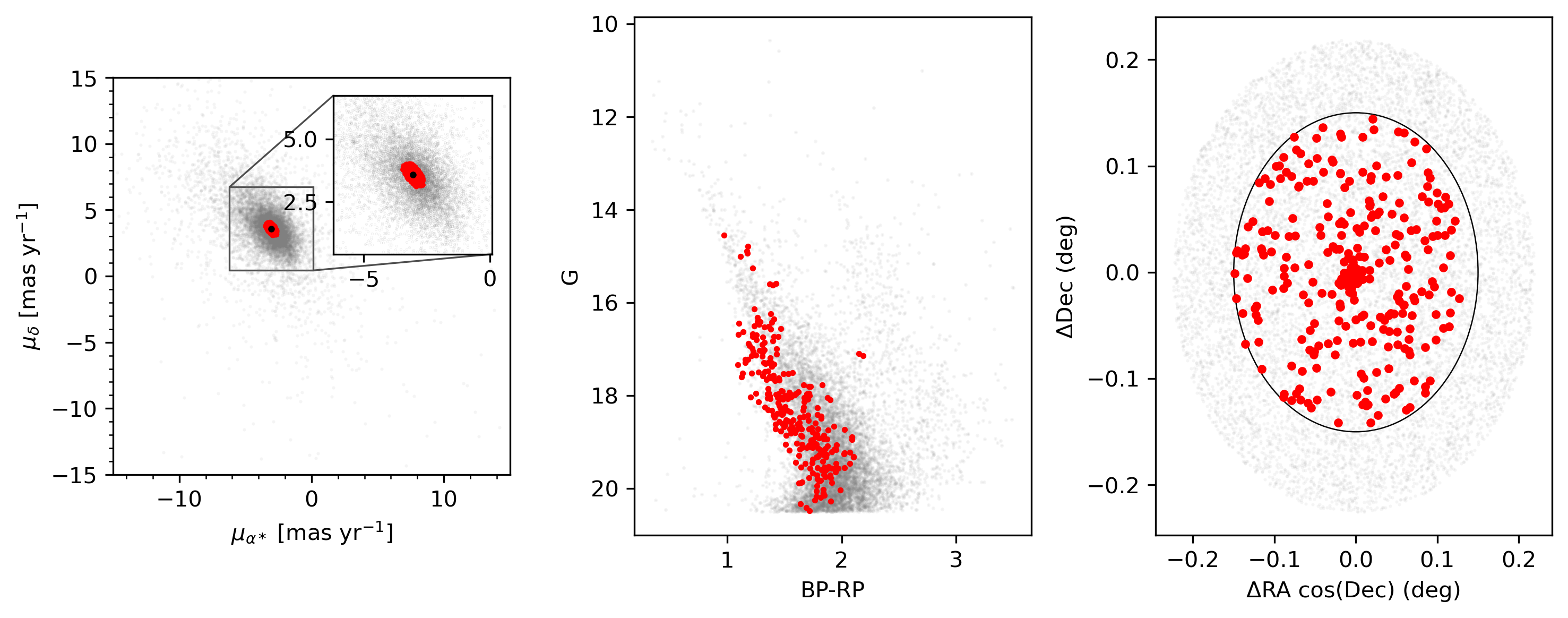}\\
    \includegraphics[width=1\linewidth]{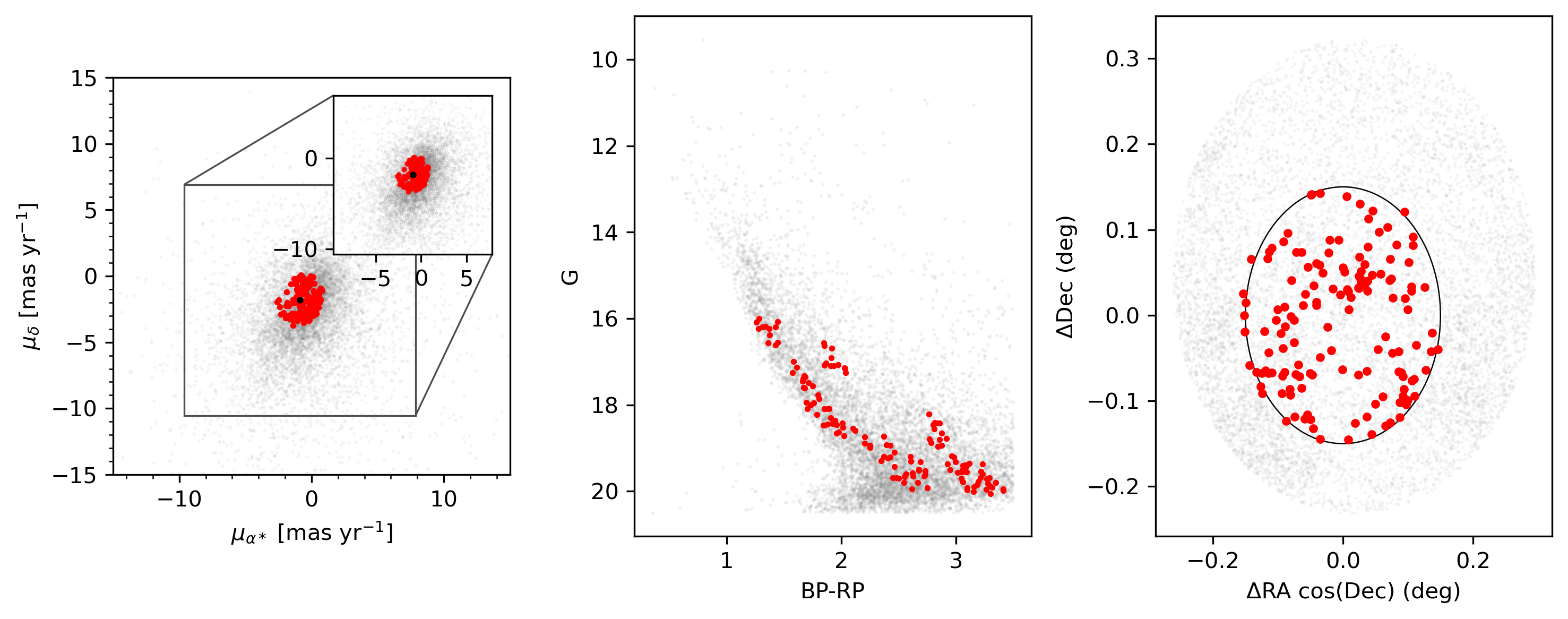}
    \includegraphics[width=1\linewidth]{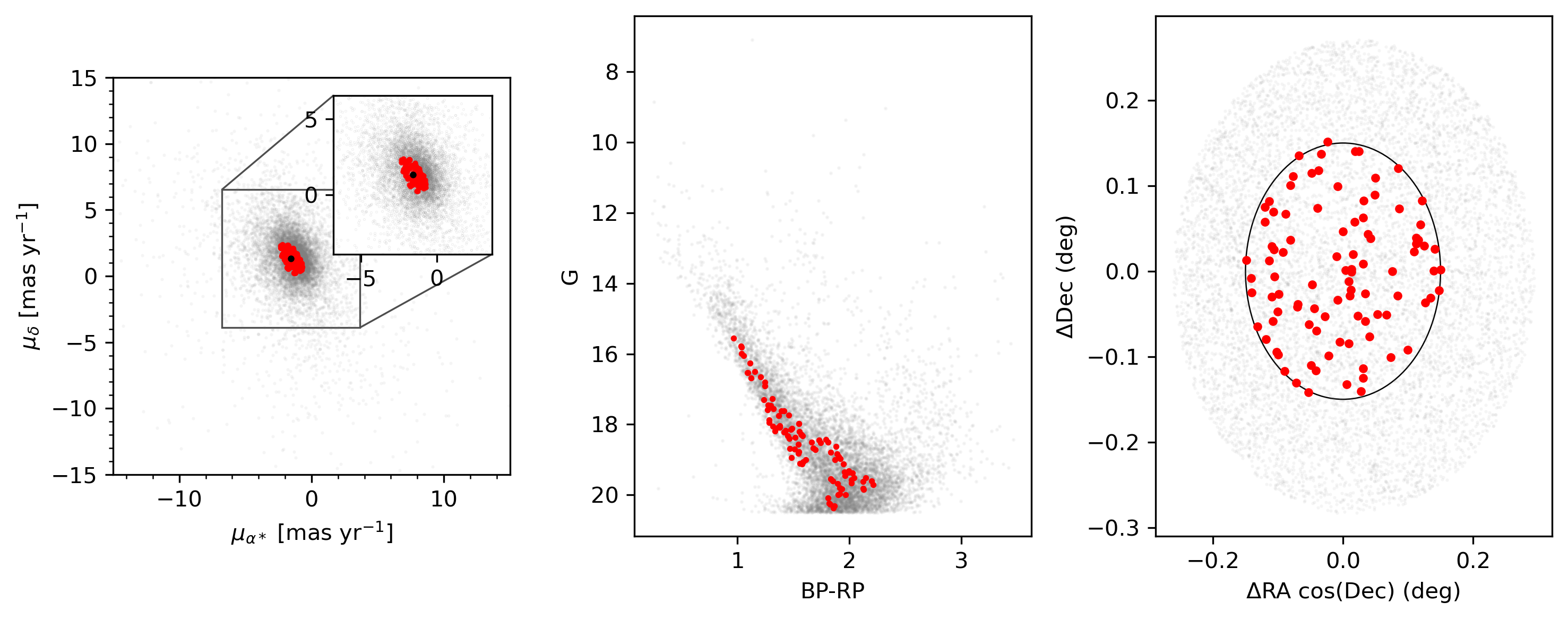}
    
    \caption{Diagnostic plots for a candidate stellar cluster identified in the VVVX survey using Gaia data.  
Left: Proper motion diagram (\(\mu_{\alpha} \cos \delta\) vs. \(\mu_{\delta}\), in mas yr\(^{-1}\)), where gray points represent all stars within 0.4 degrees and 0.5 mas of the cluster center. A red dot where the inset line is projected indicate HDBSCAN-selected members. The inset highlights stars within 2\(\sigma\) of the proper motion centroid in black dot.  
Middle: Gaia color–magnitude diagram (BP–RP vs. G), showing HDBSCAN-selected cluster members (red) tracing a distinct main sequence, albeit contaminated among field stars (gray).  
Right: Spatial distribution in \textit{\(\Delta\)RA\(\cos\)(Dec)} and \textit{\(\Delta\)Dec (degrees)}. HDBSCAN selected Cluster members (red) are concentrated within 0.3 degrees of the center, while field stars (gray) appear scattered. We indicate the cluster overdensity, centered on the black circle with a radius of 0.15${^\circ}$, which is clearly distinct from the surrounding field.  Each panel corresponds to one of the clusters shown in Figure~\ref{fig:composite}: the first panel shows the cluster centered on tile 0602 (panel a), the second panel on tile e1022 panel b, and the third and fourth panels on tiles e0965 (panel c) and e1047 (panel d), respectively.
}
    \label{fig:gaiaimage}
\end{figure}
Furthermore, we retrieved the \textit{Gaia} counterparts for each detected overdensity and analysed three distinct properties: proper motion, CMD, and spatial stellar distribution. These were used to assess whether the identified overdensities show the expected characteristics of stellar clusters. Figure~\ref{fig:gaiaimage} presents these diagnostics for the clusters shown in Figure~\ref{fig:composite}. The left panel shows the distribution of stars in proper motion space. The x-axis represents the proper motion in ($\mu_{\alpha} \cos \delta$), and the y-axis shows the proper motion in ($\mu_{\delta}$), both in (mas\,yr$^{-1}$). All stars within a radius of r $\sim$ 0.4 deg. and a parallax tolerance of 0.5 mas are plotted in gray. The cluster members, identified using the HDBSCAN algorithm, are highlighted in red, and the inset is projected onto this region. The inset in the upper-left corner zooms into the overdense area, where red points mark stars located within a $2\sigma$ threshold around the proper motion peak indicated with black dot.
This tight grouping confirms the reliability of the identified members.

The middle panel shows the \textit{Gaia} CMD, with BP$-$RP color vs G-band magnitude. gray points represent the full stellar population in the field, while red points indicate the cluster members, albeit with some residual field contamination. These members form a clear sequence that is consistent with a single-age stellar population, suggesting the presence of a main sequence and possibly a turn-off or red giant branch. In contrast, the broader distribution of field stars, particularly at fainter magnitudes ($G > 16$), highlights the effectiveness of the selection criteria based on proper motion and parallax.

The right panel shows the spatial distribution of stars around the cluster center. The x-axis indicates the offset in \textit{($\Delta\mathrm{RA} \cos \mathrm{Dec}$)} and the y-axis the offset in \textit{($\Delta\mathrm{Dec}$)}, both in degrees. Gray points represent all stars within a radius of approximately 0.4°, while red points mark the HDBSCAN-selected cluster members, which still show some residual field contamination. The members are concentrated within about 0.3° of the cluster center, consistent with a typical tidal radius, whereas the field stars are more widely scattered. The central overdensity of each cluster is also indicated by a gray circle.

\begin{figure}
    \centering
    \includegraphics[width=0.9\linewidth]{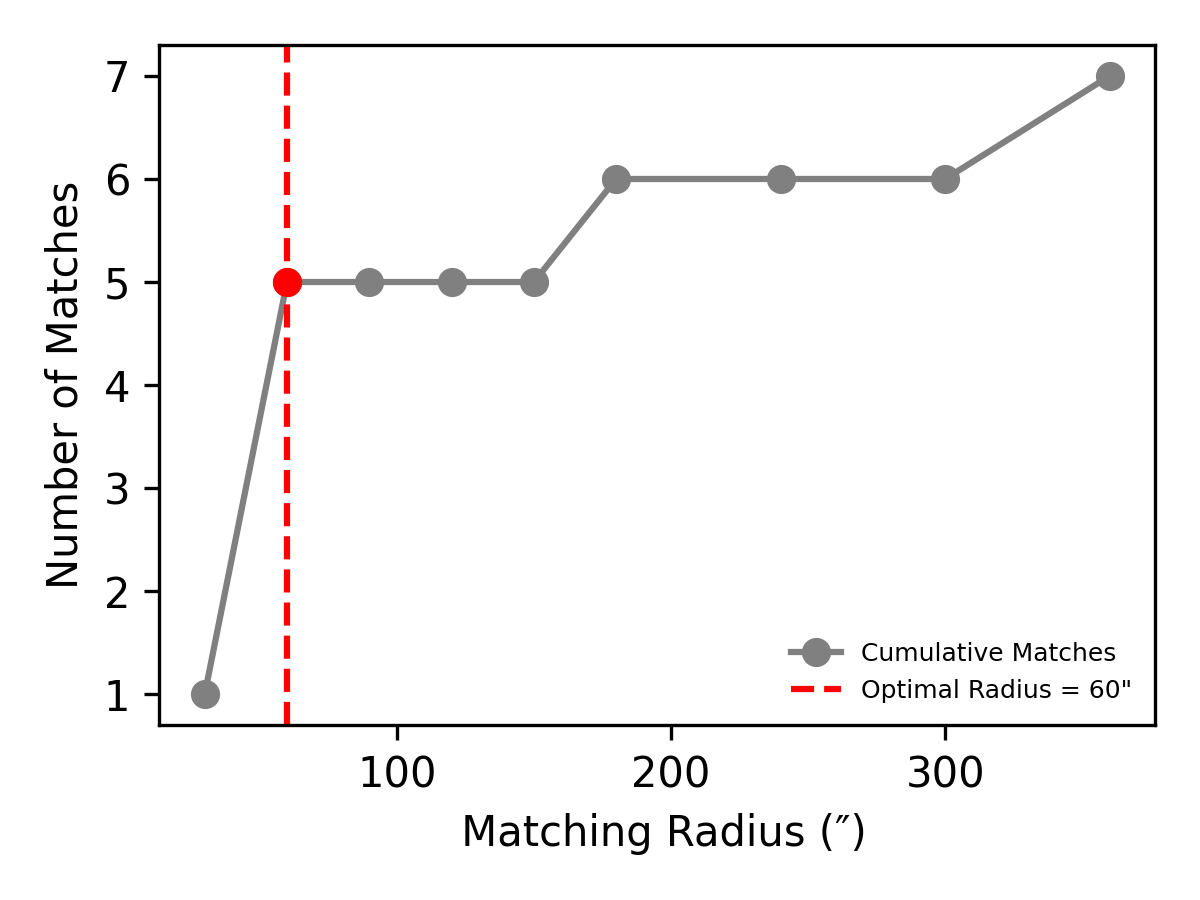}
    \caption{Cross-match analysis of 60 candidate stellar overdensities with recent cluster catalogs}
    \label{fig:match}
\end{figure}

To determine whether the remaining 45 detections are new, we cross-matched them against recently published cluster catalogs \citet{qin2023hunting, hunt2023improving, he2022unveiling, gupta2024obscured}. Given the high stellar density and source confusion in the inner bulge/disk, as well as possible astrometric offsets between catalogs, we performed a cross-match analysis to evaluate the optimal radius for source association. We tested a range of radii from 30$''$ to 360$''$ and calculated the cumulative number of matches as a function of radius. Figure~\ref{fig:match} shows the resulting curve: the number of matches increases steeply up to $\sim$60$''$, beyond which the growth plateaus, indicating the increasing contribution of spurious matches. We adopt a matching radius of 60$''$, marked by the vertical red dashed line and highlighted at the point (60$''$, 5). While this value captures most genuine associations, it is unusually large and may reflect underlying astrometric discrepancies, which will be further assessed in future work.
The coordinates of our cluster candidates used for the cross-match correspond to the centroid of the stellar overdensity identified by \textsc{CANDiSC}. For the reference catalogs, we adopted the published central coordinates. This approach ensures that the cross-match is performed consistently using centroid-based positions in both datasets.
The cross-match revealed:
\begin{itemize}
    \item 0 match with the catalog of \citet{qin2023hunting},
    \item 5 matches with \citet{hunt2023improving}, but the number increase to 7 at 360$''$
    \item 1 matches with \citet{he2022unveiling}, at 300$''$ radii,
    \item 0 matches with \citet{gupta2024obscured}.
\end{itemize}

\begin{table}[ht!]
\caption{Summary of cluster detection and cross-identification results.}
\label{tab:summary_detection}
\centering
\begin{tabular}{l c}
\hline
Category & Count \\
\hline
Total detections by \textsc{CANDiSC} & 163 \\
Matches in SIMBAD & 118 \\
Matches in recent catalogs & 5 \\
Unique matches among recent catalogs & 5 (after overlap) \\
New candidates & 40 \\
\hline
\end{tabular}
\tablefoot{Recent catalogs include those of \cite{qin2023hunting}, \cite{he2022unveiling}, \cite{hunt2023improving}, and \cite{gupta2024obscured}.}
\end{table}
After accounting for all overlaps, we identify 40 candidate clusters that, to the best of our knowledge, are not present in any existing public catalog, assuming a 60$''$ matching radius. A summary of the detection and cross-matching results is provided in Table~\ref{tab:summary_detection}. Table \ref{tab:cluster_New} and Table \ref{tab:recovered_clusters} show the 40 new candidates and 10 rows of the recovered clusters, respectively. The complete catalogs comprising both tables are available online in machine-readable format.\footnote{\url{The Link}} These new candidates will be discussed in detail in a forthcoming paper.

Figure~\ref{fig:Bical_distribu} shows the distribution of cluster candidates identified by \textsc{CANDiSC} in Galactic coordinates. 
The upper-left panel presents these candidates overlaid on an Aitoff projection of the VVVX survey footprint. Red points indicate our new detections, while blue points correspond to 788\footnote{The complete list of 788 VVV/VVVX clusters will be made available at CDS: \url{http://cdsarc.u-strasbg.fr}}
 previously known inner bulge/disk clusters compiled from literature sources based on VVV and VVVX data \citep{barba2015hundreds,barba2019sequoia,bica2018new,bidin2011three,borissova2011new,borissova2014new,borissova2018new,borissova2020small,camargo2019three,dias2022fsr,garro2020vvvx,garro2021confirmation,garro2022inspection,garro2022new,garro2022unveiling,garro2024vvvx,ivanov2017candidate,minniti2011discovery,minniti2017elephant,minniti2017new,minniti2017B,minniti2019fifty,minniti2021discoveryb,minniti2021discoveryc,minniti2021eight,obasi2021confirmation,saroon2024three}. 
The upper-right panel compares our detections with the distribution of known Milky Way clusters from the catalogue of \citet{bica2018multi}, which contains over 10,000 entries. Galactic longitude ($l$) ranges from $-60^\circ$ to $+60^\circ$, and latitude ($b$) from $-15^\circ$ to $+15^\circ$.

We also show a smoothed stellar density model, 
combining a bulge component with exponential scale lengths $l_0 = 10^\circ$ and $b_0 = 5^\circ$ and a disk component with $l_0 = 20^\circ$ and $b_0 = 2^\circ$. The density field is normalised and color-coded, with a horizontal color bar indicating the scale. Known clusters from \citet{bica2018multi} are plotted as gray points, with point sizes scaled to their major axis in arcminutes; a separate color bar indicates this scale.  
Axis labels indicate Galactic longitude and latitude in degrees, and a grid is overlaid for reference. The legend in the upper-right corner identifies all plotted elements.

The bottom panel shows the same data in equatorial coordinates (RA, Dec). Our new detections follow the overall spatial distribution of known VVV/VVVX clusters, supporting the robustness of the search method. 
At the same time, they extend the census into regions of the VVVX footprint that were sparsely populated in the literature, thereby highlighting underexplored areas of the inner Galaxy.

\begin{figure*}
    \centering
    \includegraphics[width=0.6\linewidth]{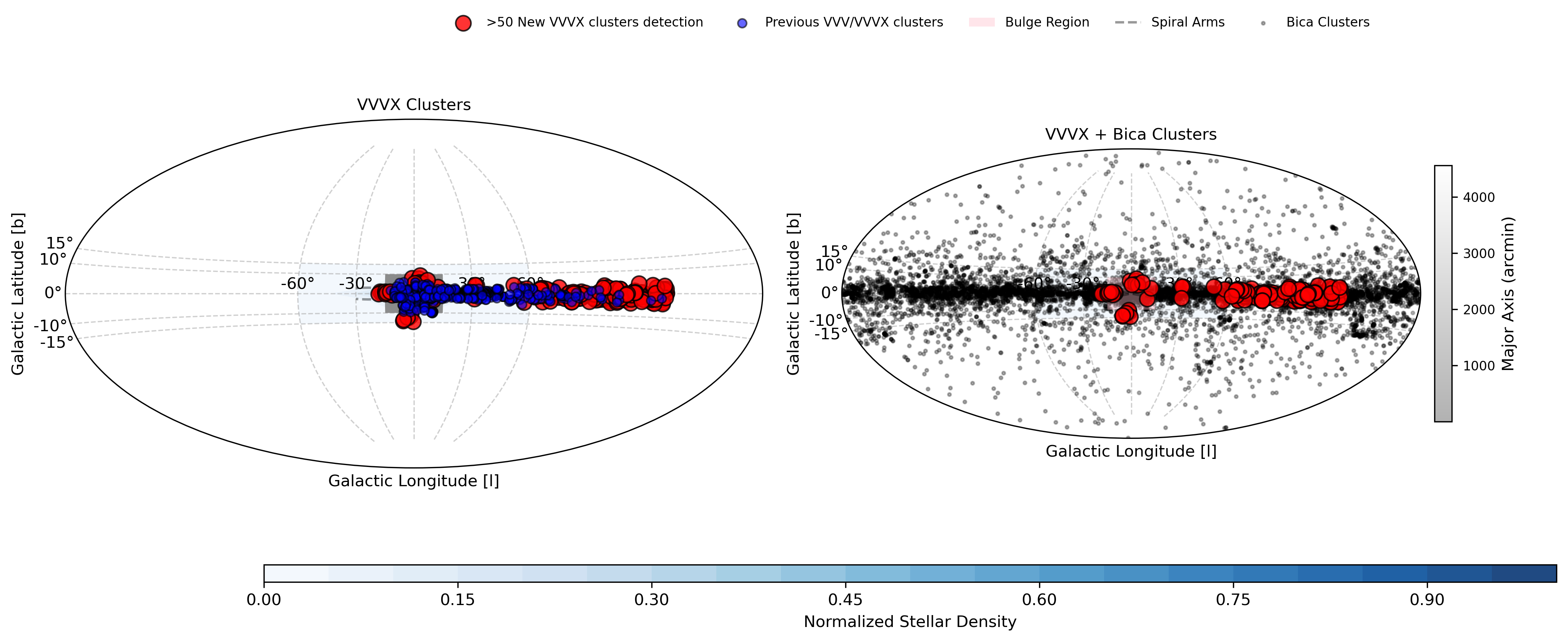}
     \includegraphics[width=0.6\linewidth]{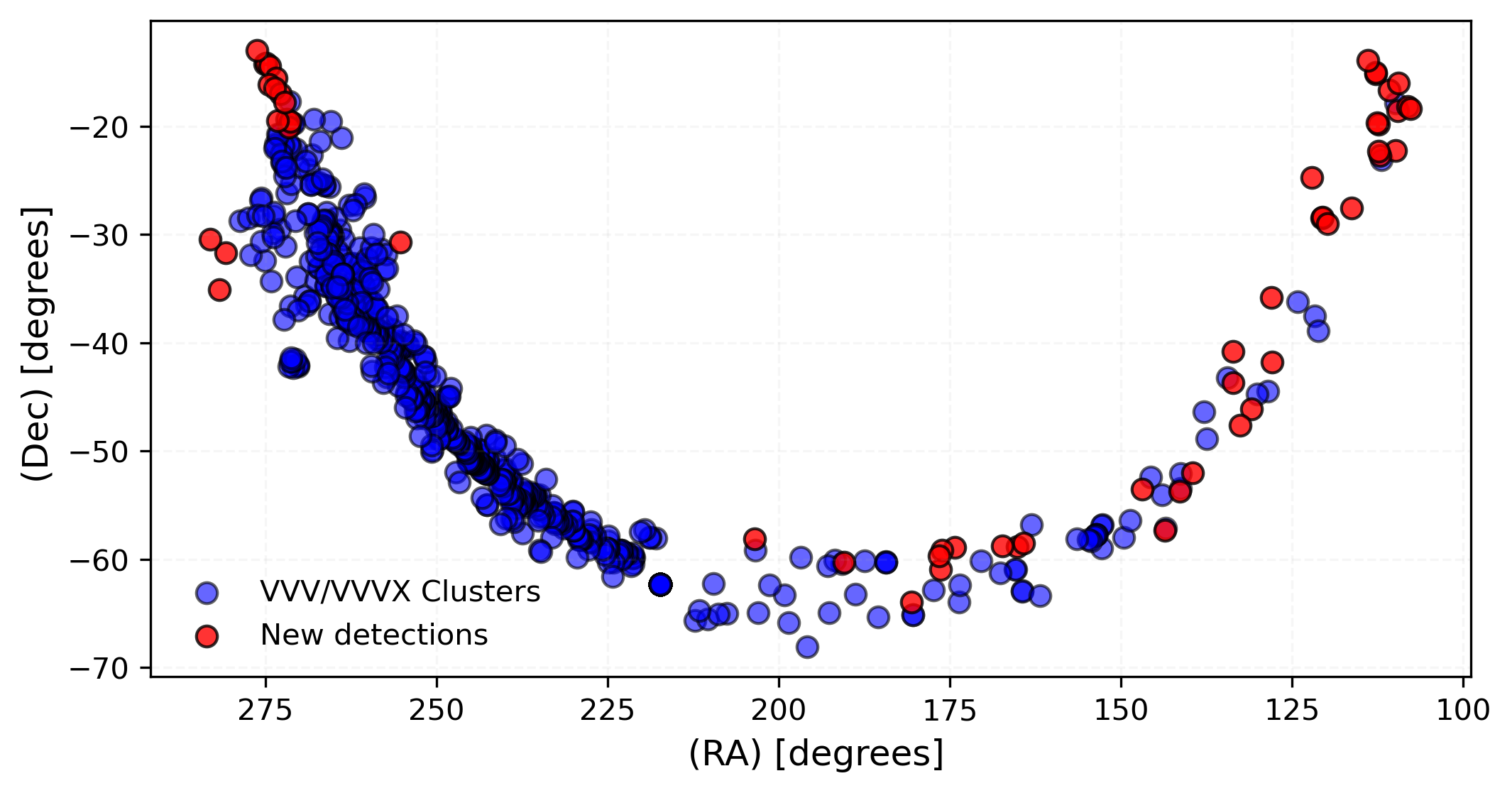}
 \caption{Left upper panel: Spatial distribution of newly detected VVVX clusters (red) overplotted with previously detected VVV/VVVX clusters (blue). 
Right upper panel: Same plot, but showing only our new detections and known clusters from \citet{bica2018multi} (gray). 
Bottom panel: Our new detections (red) overplotted with previously known VVV/VVVX clusters in equatorial coordinates (RA, Dec).
}
    \label{fig:Bical_distribu}
\end{figure*}
\subsection{Limitations of \textsc{CANDiSC}}
One of the limitations of \textsc{CANDiSC} is that the cluster membership recovery depends on the adopted color cut. In stellar clusters, especially those in the inner bulge and disk; stellar colors vary significantly due to a combination of extinction, crowding, and metallicity effects. Although stellar color broadly correlates with cluster age, this relationship is complex in these regions. The color cuts inherently limit the range of physical parameters such as age, extinction, and distance for which the cluster search is effective. This is particularly relevant for clusters in regions with high/spatial differential reddening or those at greater distances, where the stellar colors may be shifted outside the initial cut range.

After testing several color cuts, we adopted an initial selection of $(J - K_s) \in [0.4, 1.4]$, which is generally sensitive to a wide range of cluster populations. In some cases this cut yielded only a single likely member, but such detections were sufficient to motivate a refined cut for verifying potential overdensities. This choice may also explain the lack of overlap with mid-infrared clusters identified by \citet{gupta2024obscured}.

Our default selection entirely missed the most heavily reddened inner-bulge systems, such as Liller~1 \citep{pallanca2021high}, 2MASS-GC02 \citep{alonso2015variable}, and VVV-CL160 \citep{minniti2021discoveryb}, which have extreme colors ($J - K_s \sim 1.5$–4), outside the adopted range. The challenge is that no single color cut can isolate all clusters simultaneously without first correcting the tiles for differential reddening, a procedure that is computationally prohibitive to apply on the fly. While the color cut we adopted performs well for relatively less reddened regions, it limits the recovery of clusters with extreme reddening or those in the far bulge. Nonetheless, this limitation is a trade-off between computational efficiency and completeness.

Although these objects lie outside our primary search region, which is relatively less reddened and well-suited to the adopted cut, we tested their recovery using tailored ranges matched to their known reddening. For Liller~1 we adopted $1.8 < J - K_s < 2.8$, $K_s < 15$; for 2MASS-GC02, $2.0 < J - K_s < 4.0$, $K_s < 17.5$ while for VVV-CL160 we adopted, $1.6 < J - K_s < 2.4$, $K_s < 16.0$ . This approach successfully recovered both systems, but failed to retrieve VVV-CL160. 

This experiment suggests an avenue for future work: systematically applying color cuts tailored to differential reddening across the bulge may reveal additional, heavily obscured globular clusters that remain undetected.

Remarkably, even with this initial color cut, \textsc{CANDiSC} successfully identified several known clusters with very low membership counts, underscoring its sensitivity. For example, tile \texttt{e1090} at $RA, Dec = 174.30^\circ, -62.74^\circ$ yielded only two members but corresponds to [FSR2007] 1591 in the \citet{kharchenko2013global} catalog. Similarly, only two members were identified in tile \texttt{e1058} ($RA, Dec = 119.84^\circ, -30.76^\circ$), associated with [FSR2007] 1336 \citep{kharchenko2013global}. Even more striking are cases where only a single member was detected, yet these correspond to known clusters: e.g., \texttt{e0692} ($RA, Dec = 114.29^\circ, -26.37^\circ$) associated with CL Alessi 18 \citep{kronberger2006new}, \texttt{e0609} ($113.86^\circ, -27.69^\circ$) with ESO 429-3 \citep{kharchenko2013global}, and \texttt{e1095} ($113.69^\circ, -19.80^\circ$) with DSH S0734.6--1947 \citep{kronberger2006new}. Notably, the globular cluster NGC 6355 \citep{di2006rr} in tile \texttt{b0463} ($260.99^\circ, -26.35^\circ$), located in a high-extinction bulge region, was also identified by just one member.
\begin{table}[ht!]
\caption{Effect of $(J - K_s)$ Colour Cuts on Star Recovery in Known Clusters}
\label{tab:colorcuts}
\centering
\scriptsize
\setlength{\tabcolsep}{3pt} 
\begin{tabular}{@{}c l l c c ccc r@{}}
\hline
Tile Name & Literature Name & Obj$_{type}$ & RA & Dec & A & B & C & D\\
\hline
e1090  & [FSR2007]1591     & OpC & 174.30 & -62.74 & 2 & 0&1&0    \\
e1058  & [FSR2007]1336     & OpC & 119.84 & -30.76 & 2 & 8&1&0    \\
e0692  & CL ALessi     & OpC & 114.29 & -26.37& 1 & 15&0&0    \\
e0609  & ESO 429-3     & OpC & 113.86 & -27.69 & 1 & 15&0&0    \\
e1095  & DSH S0734.6--1947     & OpC & 113.69 & -19.80 & 1 & 6&6&7    \\
b0463  & NGC 6355     & GC & 260.99 & -26.35 & 1 & 0&0&5    \\

\hline
\end{tabular}
\caption*{\textbf{Note:} Columns A, B, C, and D show the total number of cluster members recovered using the following color cuts, respectively: \(0.4 \leq J - K_s \leq 1.4\), \(0.3 \leq J - K_s \leq 1.4\), \(0.6 \leq J - K_s \leq 2.0\), and \(0.5 \leq J - K_s \leq 1.4\).
 }
\end{table}
Following the identification of these clusters with low member counts, we broadened the adopted color-cut range to improve the isolation and recovery of cluster members. Because these clusters are generally redder than our default color selection, primarily due to spatial differential reddening expanding the color range, as summarized in Table~\ref{tab:colorcuts}, allowed us to retrieve additional members. Overall, the initial $(J - K_s)$ range of [0.4, 1.4] remained the most robust for detecting diverse cluster populations across the survey.

In general, after testing various color cuts, we find that when the number of recovered members is fewer than five in all the color cuts adopted, the corresponding target should be treated with caution.
\subsection{False positives}
\label{false_postive}
While the \textsc{CANDiSC} algorithm demonstrates high reliability, we assessed its susceptibility to false positive detections, particularly in cases where fewer than five members are recovered. Such low-member detections are more prone to contamination from non-cluster sources, including isolated evolved stars, variable stars, and dark nebulae that may mimic the spatial profile of genuine clusters. Out of the 163 targets identified by \textsc{CANDiSC}, we find three objects that are likely false positives based on cross-identification with SIMBAD:
\begin{itemize}
    \item Tile \texttt{e0787}, at coordinates $RA,Dec = 131.85^\circ,\ -39.03^\circ$, matches the dark nebula TGU H1669 \citep{dobashi2005atlas}, with two members detected by \textsc{CANDiSC}.
    \item Tile \texttt{e0770}, at $RA,Dec = 115.53^\circ,\ -17.82^\circ$, corresponds to the variable star ATO J115.5033$-$17.8211 \citep{zacharias2013fourth}, with one member recovered.
    \item Tile \texttt{e0854}, at $RA,Dec = 117.33^\circ,\ -19.05^\circ$, matches the eclipsing binary \textit{Gaia}~DR3~5715909087889352320 \citep{eyer2023gaia}, with three members detected.
\end{itemize}

To assess the robustness of our selection criteria against contamination, we varied the color cut for these false-positive cases by adopting the following intervals: \(0.3 \leq J - K_s \leq 1.4\), \(0.6 \leq J - K_s \leq 2.0\), and \(0.5 \leq J - K_s \leq 1.4\). For tile \texttt{e0787}, zero members were recovered across all cuts. Similarly, tile \texttt{e0770} returned zero detections for each cut. For tile \texttt{e0854}, only two members were recovered in the first interval, with none in the others. These results suggest that contaminating sources rarely reproduce the typical color distribution of real clusters, which generally spans \(J - K_s\) values from 0.3 to 2.0. This provides an additional validation step that supports the reliability of our final cluster sample.
If we consider only these three confirmed misidentifications, the contamination rate is $3/163 \approx 1.8\%$. We note that this value represents only confirmed misidentifications based on visual and photometric inspection, and therefore should be considered a lower limit to the true contamination rate, as additional false positives may remain among the unconfirmed candidates. 

We emphasize that all candidate detections were visually inspected using composite color images (PanSTARRS \citep{tonry2012pan} and DECAPS DR1 \citep{schlafly2018decam}).

\subsection{Parameter Sensitivity}
We also assessed the robustness of the \textsc{CANDiSC} detections by performing an internal validation by varying the core parameters of the algorithm: the KDE bandwidth $h$, the DBSCAN radius $\varepsilon$, the number of neighbours $k$ for NNDE, and the detection threshold $\sigma$. These parameters were adjusted independently within plausible ranges informed by visual inspection and empirical tuning (see Sect.\ref{sec:validation}). We find that the number of detected overdensities remains relatively stable within the range $h = 0.08^\circ$–$0.12^\circ$, $\varepsilon = 0.08^\circ$–$0.15^\circ$, and $k = 3$–$6$. Beyond these ranges, the algorithm either begins to over-smooth the stellar density field, merging distinct structures, or becomes overly sensitive to noise, particularly in sparse fields.

\subsection{Statistical Summary of Detections}

In total, \textsc{CANDiSC} identified 163 candidate stellar overdensities across the studied 680 VVVX tiles. The number of cluster members per detection spans a wide range, from 1 up to more than 1000. Figure~\ref{fig:hist_members} presents the distribution of member counts for all detected clusters. Most detections fall within the range of 5 to 100 members, with a peak near 12. A substantial number of candidates have between 12 and 40 members, while a smaller group exceeds 100 members. Detections with fewer than five members were reviewed individually, as discussed in Sect.~\ref{false_postive}. This assessment is based on the default color cut; varying the color cut thresholds can affect the number of recovered members, as previously discussed.

\begin{figure}[ht!]
\centering
\includegraphics[width=\linewidth]{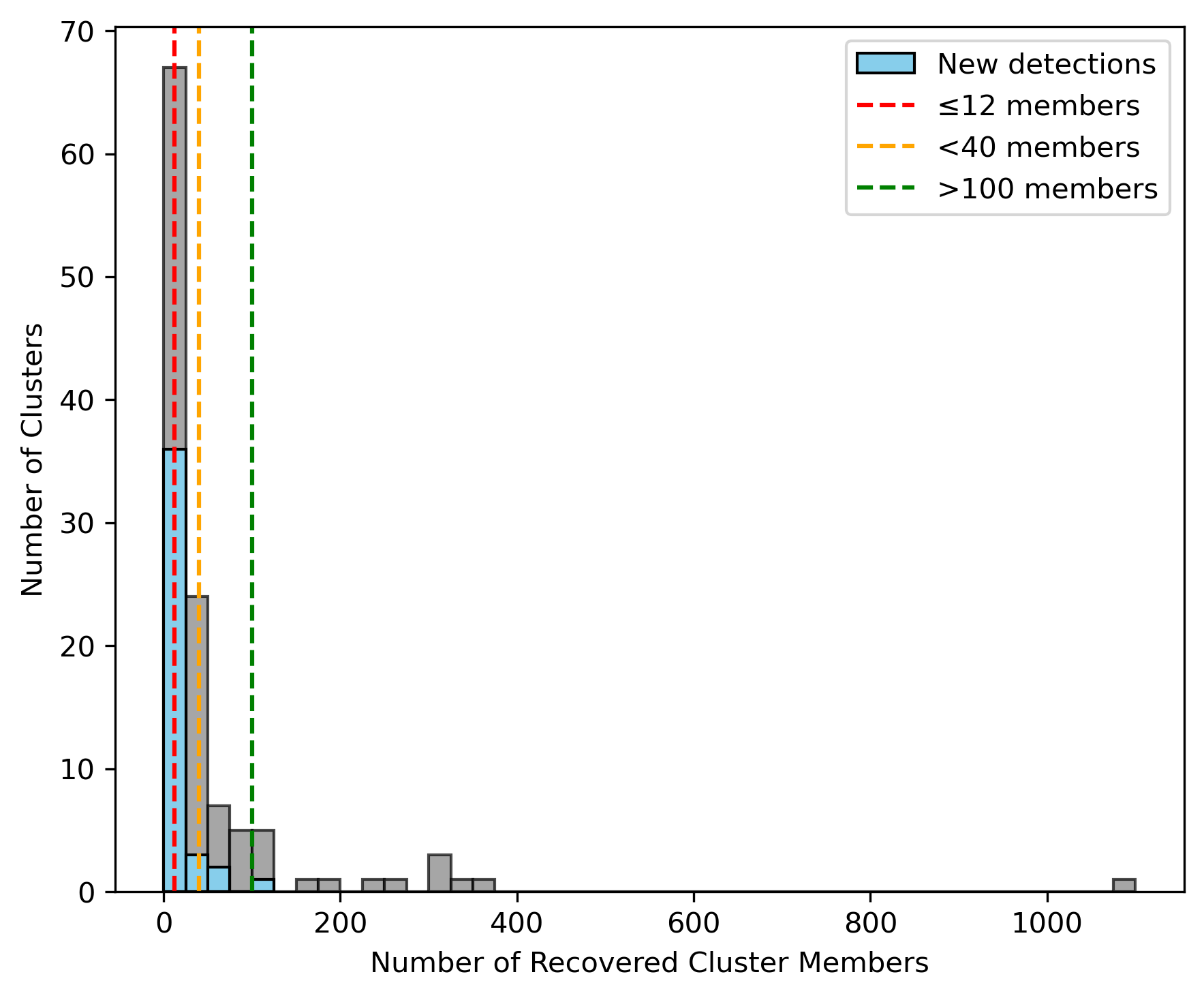}
\caption{Distribution of recovered cluster members across all 163 detected overdensities. The blue shade shows new detection in this work.}
\label{fig:hist_members}
\end{figure}

\section{Discussion }
\label{section5}
The \textsc{CANDiSC} algorithm introduces a consensus-based, unsupervised approach for the detection of stellar clusters, optimised for wide-field photometric surveys. Designed initially for the VVVX survey, \textsc{CANDiSC} operates on a minimal set of inputs, namely stellar coordinates and a photometric color-magnitude filter, and requires no prior assumptions about cluster morphology or location. Its architecture combines three independent density-based clustering methods (KDE, DBSCAN, NNDE), flagging an overdensity only when detected by at least two methods. This strategy enhances robustness and reduces false positives while preserving sensitivity to a wide range of cluster morphologies.

Unlike earlier efforts that relied heavily on manual inspection of individual tiles to identify overdensities \citep[e.g.,][]{bica2018new,garro2020vvvx,garro2022unveiling, garro2024vvvx,minniti2017new,minniti2021discovery,minniti2021new}, \textsc{CANDiSC} enables a fully automated, homogeneous analysis of the entire VVVX footprint. When applied to all 680 VVVX tiles, \textsc{CANDiSC} successfully recovered all previously known clusters in the region and identified several new candidates, including objects likely missed due to their low surface density or high extinction. This highlights the algorithm’s potential to uncover hidden stellar systems that escape visual detection.

While these results are promising, several limitations must be acknowledged. First, the consensus requirement can lead to missed detections: if an overdensity is detected by only one of the three methods, it is discarded. This deliberate design choice prioritizes purity over completeness, aiming to limit contamination. As shown in Sect.~\ref{false_postive}, the lower limit for the contamination rate is estimated to be 1.8\%, primarily due to isolated evolved stars or compact nebular structures misidentified as clusters. Second, due to hardware constraints, a downsampling step was implemented for tiles containing more than $\sim$1.2 million stars (see Sect.\ref{data_processing}). This step may introduce incompleteness in particularly crowded or diffuse regions. However, stress tests across multiple downsampling levels (10\%, 20\%, 50\%, 80\%, and 90\%) confirmed that \textsc{CANDiSC} consistently recovered known clusters, even when visual overdensities were no longer apparent. We therefore consider this limitation to be well-controlled in practice.

Despite these caveats, the performance of \textsc{CANDiSC} validates its utility as a scalable and objective tool for cluster detection in large photometric surveys. Its sensitivity to low-density structures and its ability to suppress false positives make it well suited for applications beyond VVVX, including forthcoming datasets from \textit{LSST} \citep[LSST collaboration][]{abell2009lsst} and \textit{Euclid} \citep{blanchard2020euclid}. A follow-up study (Obasi et al., in prep.) is currently underway to characterise the new cluster candidates using \textit{Gaia}~DR3 \citep{brown2021gaia}, 2MASS \citep{skrutskie2006two}, and Dark Energy Camera Plane Survey \citep[DECaPS;][]{schlafly2018decam,saydjari2023dark}. This will enable the derivation of physical and structural parameters such as extinction, distance, metallicity, and age. The study will provide further insight into the formation and dynamical evolution of these systems within the Galaxy.

\section{Final Remarks}
\label{section6}	
We have developed and applied \textsc{CANDiSC}, a Consensus-based Algorithm for Nonparametric Detection of Star Clusters algorithm for the detection of stellar clusters, to the full VVVX survey footprint shown in Figure \ref{fig:survey_area}. Our method combines three independent clustering techniques (KDE, DBSCAN, NNDE) and flags a stellar overdensity only when detected by at least two of these techniques independently.

Applying \textsc{CANDiSC} to 680 VVVX tiles, we identified 163 cluster candidates. Of these, 118 match known clusters in the SIMBAD database, while 5 correspond to entries in recently published catalogs not yet reflected in SIMBAD. We find 40 candidates that appear to be previously uncatalogued and may represent new stellar clusters. These include objects located in regions of high extinction and/or low surface density, which are often missed by traditional methods.

The success of \textsc{CANDiSC} in recovering all previously known clusters in the VVVX footprint, while also uncovering new systems, demonstrates its robustness and reliability. It provides a valuable tool for mining large photometric datasets in a reproducible and scalable manner. In a forthcoming paper, we will characterise these new candidates using complementary datasets including \textit{Gaia} DR3, 2MASS, and DECaPS and VVVX, to derive distances, extinctions, metallicities, and ages.

\begin{appendix}

  \section{Detected clusters} 
  
We show in Table~\ref{tab:cluster_New} the 40 rows of new cluster candidates detected by \textsc{CANDiSC} that are not currently listed in the SIMBAD database. Column 1 contains the VVVX cluster IDs, Column 2 the corresponding VVVX tile names, and Columns 3 and 4 the equatorial coordinates (RA, Dec). Columns 5–8 present the number of sources recovered for different color cuts, as detailed in the Table \ref{tab:colorcuts} footnote. 
Table~\ref{tab:recovered_clusters} presents the first 10 rows of previously known clusters recovered by \textsc{CANDiSC}, including their literature names, tile names, and coordinates (RA, Dec). The complete catalog of recovered clusters is also provided as a machine-readable table.
\begin{table*}
\centering
\small
\caption{New cluster candidates detected by the \textsc{CANDiSC} algorithm in the VVVX survey. }

\label{tab:cluster_New}
\begin{tabular}{l l r r r r r r}
\hline
VVVX cluster IDs & Tile name & RA & Dec & A & B & C & D \\
\hline
VVVX-Obasi 1  & e0961 & 271.556 & -20.054 & 68 & 68 & 0  & 62 \\
VVVX-Obasi 2  & e0961 & 271.902 & -19.371 & 104 & 102 & 0 & 97 \\
VVVX-Obasi 3  & e0961 & 271.353 & -19.565 & 11 & 11 & 0 & 12 \\
VVVX-Obasi 4  & e0965 & 275.125 & -14.232 & 44 & 41 & 0 & 41 \\
VVVX-Obasi 5 & e1030 & 139.490 & -51.024 & 31 & 32 & 84 & 14 \\
VVVX-Obasi 6 & e1022 & 127.917 & -41.786 & 18 & 14 & 14 & 2 \\
VVVX-Obasi 7 & e0602 & 108.096 & -18.160 & 12 & 9 & 0 & 0 \\
VVVX-Obasi 8 & e0605 & 109.818 & -22.231 & 36 & 35 & 11 & 28 \\
VVVX-Obasi 9 & e1030 & 139.497 & -51.028 & 23 & 23 & 18 & 22 \\
VVVX-Obasi 10 & e0689 & 112.097 & -22.678 & 11 & 9 & 1 & 0 \\
VVVX-Obasi 11 & e0689 & 112.098 & -22.674 & 11 & 9 & 1 & 0 \\
VVVX-Obasi 12 & e1135 & 176.346 & -60.941 & 14 & 12 & 0 & 14 \\
VVVX-Obasi 13 & b0473 & 255.324 & -30.687 & 4 & 0 & 2 & 0 \\
VVVX-Obasi 14 & e1135 & 176.333 & -60.941 & 9 & 24 & 1 & 0 \\
VVVX-Obasi 15 & e1131 & 165.118 & -58.834 & 10 & 11 & 17 & 11 \\
VVVX-Obasi 16 & e1047 & 110.755 & -16.673 & 10 & 15 & 18 & 20 \\
VVVX-Obasi 17 & b0407 & 281.761 & -35.136 & 2 & 0 & 4 & 0 \\
VVVX-Obasi 18 & e0971 & 273.415 & -15.534 & 2 & 2 & 0 & 2 \\
VVVX-Obasi 19 & e1124 & 146.851 & -53.504 & 8 & 3 & 22 & 1 \\
VVVX-Obasi 20 & e0816 & 191.084 & -60.095 & 1 & 1 & 0 & 0 \\
VVVX-Obasi 21 & e1001 & 109.534 & -16.720 & 1 & 3 & 0 & 0 \\
VVVX-Obasi 22 & e0957 & 274.496 & -16.137 & 3 & 3 & 0 & 3 \\
VVVX-Obasi 23 & e1115 & 133.533 & -43.693 & 6 & 6 & 0 & 5 \\
VVVX-Obasi 24 & e0810 & 174.308 & -58.891 & 6 & 5 & 6 & 8 \\
VVVX-Obasi 25 & e0954 & 273.271 & -19.501 & 4 & 2 & 0 & 0 \\
VVVX-Obasi 26 & e1010 & 116.175 & -28.378 & 3 & 3 & 4 & 0 \\
VVVX-Obasi 27 & e0634 & 143.489 & -57.313 & 7 & 8 & 0 & 9 \\
VVVX-Obasi 28 & e0904 & 203.571 & -58.113 & 4 & 4 & 0 & 0 \\
VVVX-Obasi 29 & e0811 & 175.023 & -58.656 & 4 & 1 & 0 & 1 \\
VVVX-Obasi 30 & e0784 & 127.95 & -35.82 & 1 & 0 & 0 & 0 \\
VVVX-Obasi 31 & e0963 & 273.618 & -16.465 & 5 & 5 & 0 & 5 \\
VVVX-Obasi 32 & e0963 & 272.850 & -16.942 & 1 & 1 & 0 & 1 \\
VVVX-Obasi 33 & e0789 & 133.954 & -41.005 & 7 & 8 & 3 & 7 \\
VVVX-Obasi 34 & e1180 & 176.489 & -59.701 & 2 & 0 & 2 & 0 \\
VVVX-Obasi 35 & e0859 & 122.030 & -31.687 & 3 & 0 & 2 & 3 \\
VVVX-Obasi 36 & b0437 & 280.754 & -31.687 & 6 & 0 & 0 & 0 \\
VVVX-Obasi 37 & e0962 & 272.161 & -17.796 & 7 & 7 & 0 & 4 \\
VVVX-Obasi 38 & b0425 & 283.056 & -30.407 & 9 & 0 & 0 & 0 \\
VVVX-Obasi 39 & e0767 & 113.843 & -13.908 & 8 & 9 & 0 & 7 \\
VVVX-Obasi 40 & e0647 & 180.507 & -65.9881 & 2 & 4 & 0 & 1 \\
\hline
\end{tabular}
\end{table*}

\begin{table*}
\centering
\caption{Previously known stellar clusters recovered by the \textsc{CANDiSC} algorithm in the VVVX survey.  First 10 rows are shown. }
\label{tab:recovered_clusters}
\begin{tabular}{lcccr}
\toprule
Literature Name & Tile Name & RA (deg) & Dec (deg) \\
\midrule
CL Haffner 180 & e1100 & 118.167 & -26.386 \\
CL Haffner 19  & e1100 & 118.194 & -26.275 \\
M 70           & b0423 & 280.800 & -32.290 \\
NGC 6380       & e0683 & 263.620 & -39.070 \\
NGC 6256       & e0930 & 254.880 & -37.120 \\
DSH J0718.4-1734 & e0685 & 109.620 & -17.570 \\
$[FSR2007]$ 0053 & e0965 & 274.680 & -13.170 \\
M 54           & b0411 & 283.761 & -30.477 \\
NGC 6652       & b0436 & 278.940 & -32.989 \\
Cl Pismis 2    & e0621 & 124.478 & -41.679 \\
\bottomrule
\end{tabular}
\end{table*}
\section{validation}\label{AppendixB}
This Appendix provides the full set of validation diagnostics derived from the synthetic cluster injections described in Section~\ref{sec:validation}. These results expand upon the global completeness trends discussed in the main text by presenting the figure–by–figure behavior of the pipeline across the three-dimensional parameter space spanned by the simulations. The figures include recovery fraction distributions, the dependence of completeness on extinction, richness, and size, astrometric offsets, and the purity–completeness relation. Together, they provide a detailed view of the strengths and limitations of both the default and tuned configurations of \textsc{CANDiSC}.

Figure~\ref{fig:recovery} shows the histogram of recovery fractions for the synthetic VVVX-like injections (315 realizations with cluster richness $N = \{80, 100, 200, 300, 400, 500, 600\}$, half-light radius $r_h = \{0.005^\circ, 0.01^\circ, 0.03^\circ\}$, and extinction $A_V = \{0.5, 1.5, 3.0\}$~mag). The left panel (default configuration) reveals a right-skewed distribution with a dominant peak near 1.0 (perfect recovery) and a tail between 0.1 and 0.8. Approximately 32\% of clusters are perfectly recovered, primarily low-extinction, high-richness cases. In contrast, the tuned configuration exhibits denser mid-range bins (0.4--0.8), indicating fewer partial recoveries and improved handling of marginal overdensities.

Recovery efficiency declines as extinction increases. For the default configuration, completeness decreases from 0.45 at $A_V = 0.5$~mag to 0.20 at $A_V = 3.0$~mag (Figure~\ref{fig:recovery_AV}, left panel). In the tuned configuration (Figure~\ref{fig:recovery_AV}, right panel), the decline is slightly shallower, from 0.40 to 0.25 over the same interval, reflecting improved sensitivity to reddened clusters.

Figure~\ref{fig:recovery_RH} shows recovery as a function of cluster size. In the default configuration (upper left), completeness increases from approximately 0.10 at $R = 80$ to 0.40 at $R = 600$, with a decrease from $\sim$0.55 at $r_h = 0.005^\circ$ to $\sim$0.10 at $r_h = 0.03^\circ$, indicating strong performance for compact clusters but difficulty with diffuse ones. The tuned configuration (lower right) shows gains across the parameter space, rising from $\sim$0.20 at $R = 80$ to $\sim$0.45 at $R = 600$, while maintaining the expected decline with increasing $r_h$.

Astrometric accuracy remains excellent in both configurations (Figure~\ref{fig:offset}). The default setup peaks at offsets below 0.025~arcsec with a tail to 0.175~arcsec, while the tuned configuration yields a slightly tighter distribution with a shorter tail ($\sim$0.15~arcsec). The purity–completeness relation (Figure~\ref{fig:purity}) displays the typical trade-off: high purity at low completeness and declining purity at higher completeness. The tuned configuration shifts the curve marginally toward higher completeness at fixed purity.

Across the 315 realizations, the default configuration recovers 245 clusters (77.78\%), while the tuned setup recovers 226 (71.75\%). Although the tuned configuration detects fewer clusters overall, it performs better on low-richness ($N \lesssim 200$) and extended ($r_h \gtrsim 0.01^\circ$) systems, reflecting a shift from maximizing detections to improving sensitivity to faint structures.

For the main catalog, we adopt the default \textsc{CANDiSC} settings, which maximise the number of reliably recovered clusters and perform robustly for compact and moderately extincted systems. Despite slightly lower sensitivity to faint or diffuse clusters, the default configuration maintains strong astrometric precision, a low contamination rate (mean spurious fraction of 5.9\%), and a completeness of $\sim$30--35\% in typical VVVX conditions ($A_V \sim 1$--2~mag, $r_h \sim 0.01^\circ$). Incorporating multi-band priors and astrometric information (proper motion, parallax, and radial velocity), as well as applying de-noising techniques, is expected to improve the detection/accurate recovery of reddened and diffuse clusters in future work. The recovery behavior versus cluster size and extinction is well-characterized, providing a robust foundation for further tuning and follow-up studies targeting fainter or more extended cluster populations.

\begin{figure*}
    \centering
    \includegraphics[width=0.4\linewidth]{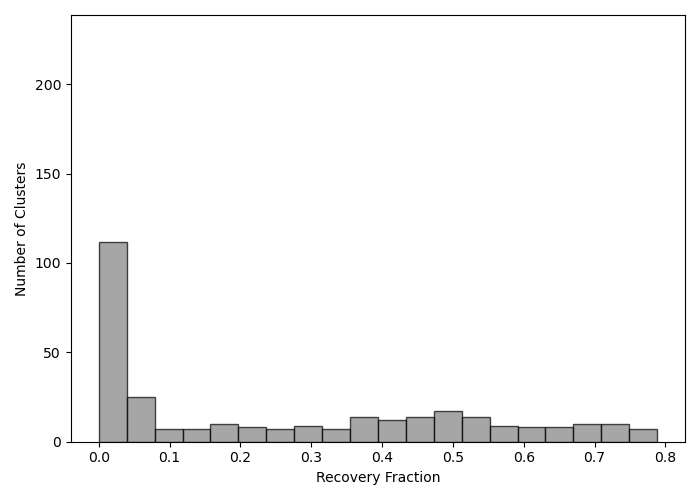}
    \includegraphics[width=0.4\linewidth]{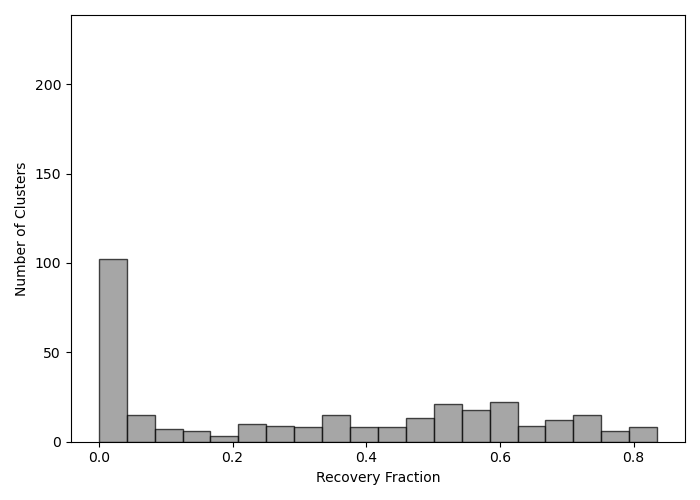}
    \caption{Histograms of recovery fractions for all injected synthetic clusters are presented for both the default and tuned configurations in the left and right panels, respectively. The distribution is right-skewed, indicating that the detection pipeline well recovers most clusters, while a small subset shows partial or failed recovery.}
    \label{fig:recovery}
\end{figure*}
\begin{figure*}
    \centering
    \includegraphics[width=0.4\linewidth]{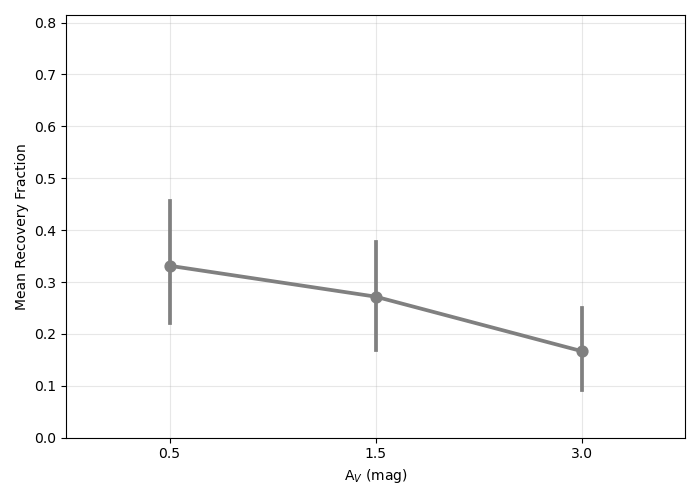}
    \includegraphics[width=0.4\linewidth]{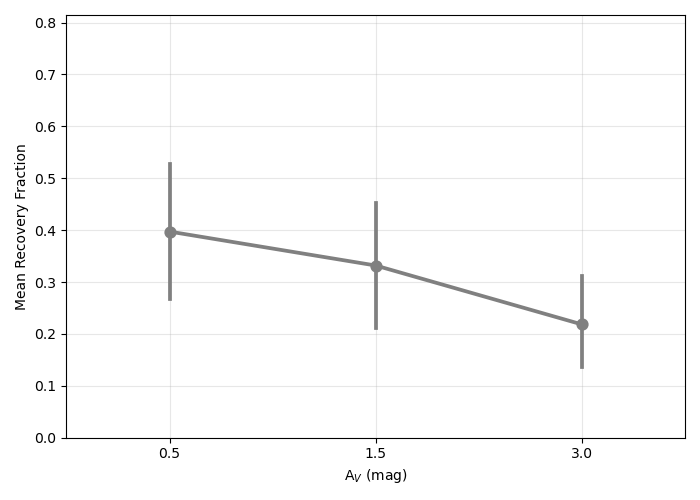}
    \caption{Mean recovery fraction as a function of extinction ($A_V$) is shown for both the default and tuned configurations. The recovery efficiency decreases systematically with increasing $A_V$, reflecting the reduced detectability of clusters toward the bulge/disk, where spatial extinction is problematic and can blend the clusters' overdensity.}
    \label{fig:recovery_AV}
\end{figure*}
\begin{figure*}
    \centering
    \includegraphics[width=0.65\linewidth]{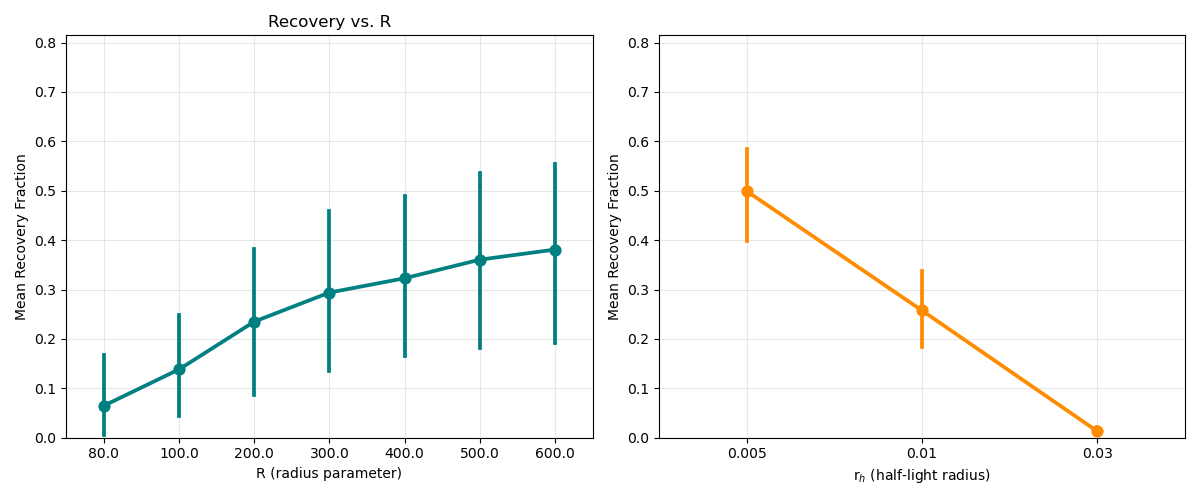}
    \includegraphics[width=0.65\linewidth]{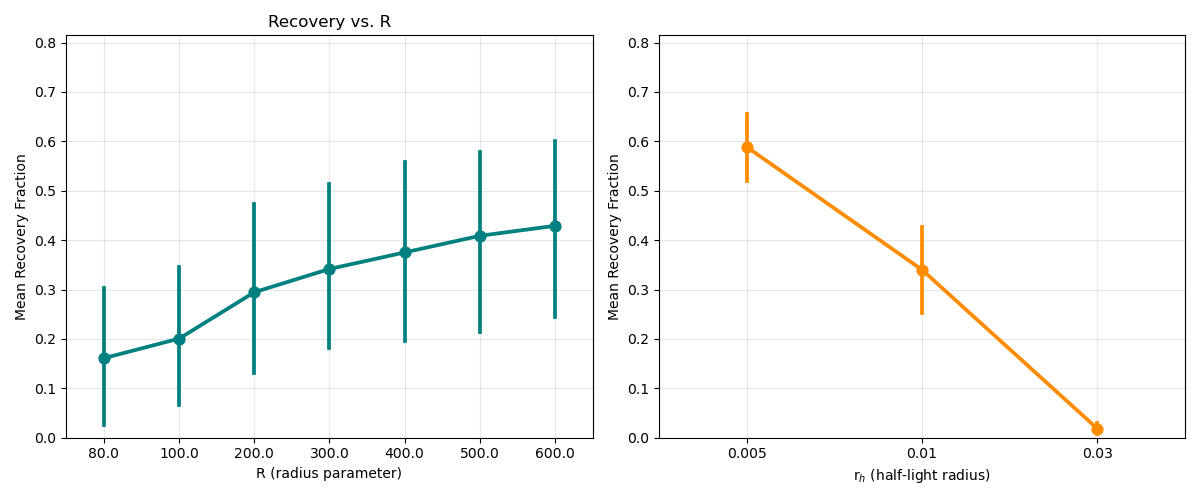}
    \caption{Mean recovery fraction as a function of the input cluster radius parameter ($R$) is shown for the default configuration in the left upper panel, while dependence of the mean recovery fraction on the half-light radius ($r_h$) is shown in the right upper panel. The bottom panel shows the same plot but for the tuned configuration.}
    \label{fig:recovery_RH}
\end{figure*}

\begin{figure*}
    \centering
    \includegraphics[width=0.4\linewidth]{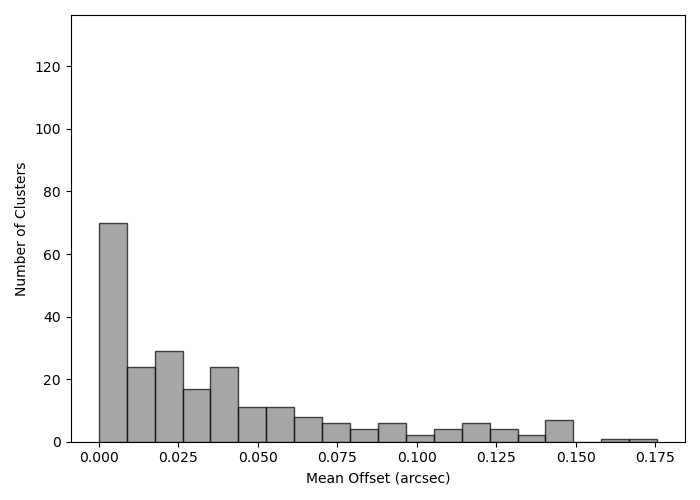}
    \includegraphics[width=0.4\linewidth]{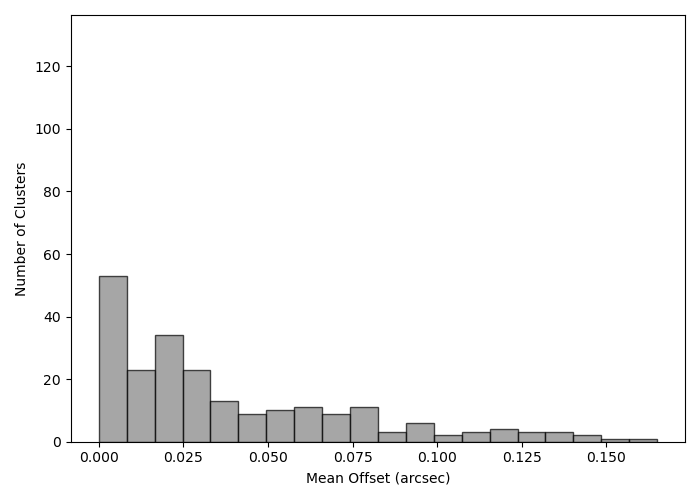}
    \caption{Histograms of the mean positional offset between injected and recovered cluster centers are shown for both default (left panel) and tuned (right panel) configurations. The distribution peaks near zero, indicating accurate centroid recovery, with a small tail toward larger offsets corresponding to marginal or blended detections.}
    \label{fig:offset}
\end{figure*}
\begin{figure*}
    \centering
    \includegraphics[width=0.4\linewidth]{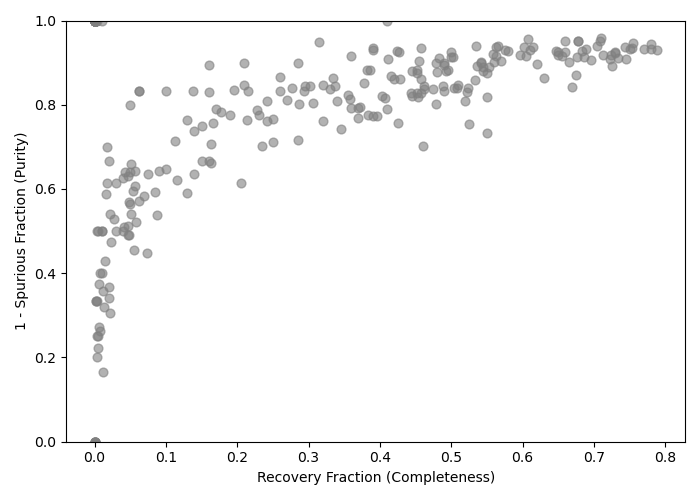}
    \includegraphics[width=0.4\linewidth]{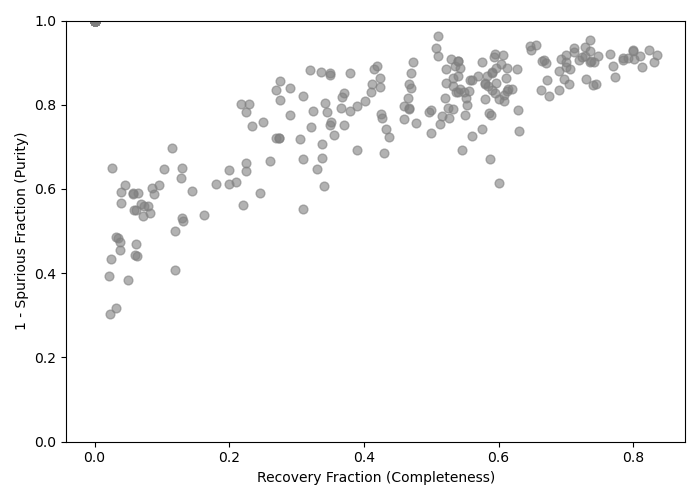}
    \caption{Scatter plot showing the trade-off between completeness (recovery fraction) and purity (1 - spurious fraction) for all injected clusters, comparing the default (left) and tuned (right) configurations. The relationship illustrates the balance between maximizing true detections and minimizing false positives.}
    \label{fig:purity}
\end{figure*}

\end{appendix}

\begin{acknowledgements} 
This work was funded by the Postdoctoral Talent Attraction Competition for Research Centers and Institutes of the Universidad Andrés Bello (UNAB) 2025, project Nº. DI-07-25/ATP. J.G.F-T gratefully acknowledges the grants support provided by ANID Fondecyt Postdoc No. 3230001 (Sponsoring researcher), the Joint Committee ESO-Government of Chile under the agreement 2023 ORP 062/2023, and the support of the Doctoral Program in Artificial Intelligence, DISC-UCN. D.M. gratefully acknowledges support from the Center for Astrophysics and Associated Technologies CATA by the ANID BASAL projects ACE210002 and FB210003, by Fondecyt Project No. 1220724. M.G. gratefully acknowledges support from Fondecyt through grant 1240755. BPLF acknowledges financial support from Conselho Nacional de Desenvolvimento Científico e Tecnológico (CNPq, Brazil; procs. 140642/2021-8 and 314718/2025-7) and Coordenação de Aperfeiçoamento de Pessoal de Nível Superior (CAPES, Brazil; Finance Code 001; proc. 88887.935756/2024-00). This study was financed, in part, by the São Paulo Research Foundation (FAPESP), Brazil; Process Number 2025/05050-3.
. J.A.-G. acknowledges support from DGI-UAntof and Mineduc-UA Cod. 2355. B.T. gratefully acknowledges support from the National Natural Science Foundation of China through grants NOs. 12473035 and 12233013, China Manned Space Project under grant NO. CMS-CSST-2025-A13 and CMS-CSST-2021-A08. R.K.S. acknowledges support from CNPq/Brazil through projects 308298/2022-5 and 421034/2023-8

  We gratefully acknowledge the use of data from the ESO Public Survey program IDs 179.B-2002 and 198.B-2004 taken with the VISTA telescope and data products from the Cambridge Astronomical Survey Unit.
\end{acknowledgements}

\bibliographystyle{aa}
\bibliography{references}

\end{document}